\documentstyle[preprint,aps,epsfig]{revtex}

\begin{document}

% \tighten
\draft
\preprint{
\vbox{
\hbox{August 1995}
\hbox{TUM/T39--95--11}
}}

\title{Nucleon Structure Functions at Moderate $Q^2$:
       Relativistic Constituent Quarks and Spectator Mass Spectrum}
\author{S.A.Kulagin$^a$, W.Melnitchouk$^b$\footnote{Present address:
	Department of Physics, University of Maryland, College Park,
	MD 20742}, T.Weigl$^b$, W.Weise$^b$}
\address{$^a$ Institute for Nuclear Research,
         Russian Academy of Sciences,
         Moscow, Russia}
\address{$^b$ Physik Department,
         Technische Universit\"{a}t M\"unchen,
         D-85747 Garching, Germany.}

\maketitle

\begin{abstract}
We present a model description of the nucleon valence structure
function applicable over the entire region of the Bjorken variable
$x$, and above moderate values of $Q^2$ ($\sim 1$ GeV$^2$).
We stress the importance of describing the complete spectrum of
intermediate states which are spectator to the deep-inelastic
collision.
At a scale of 1 GeV$^2$ the relevant degrees of freedom are
constituent quarks and pions.
The large-$x$ region is then described in terms of scattering
from constituent quarks in the nucleon, while the dressing of
constituent quarks by pions plays an important role at
intermediate $x$ values.
The correct small-$x$ behavior, which is necessary for the proper
normalization of the valence distributions, is guaranteed by modeling
the asymptotic spectator mass spectrum according to Regge
phenomenology.
\end{abstract}

\pacs{PACS numbers: 12.40.Aa, 13.60.Hb, 25.30.Fj        \\ \\ \\
      Work supported in part by BMBF.\\
      To appear in {\em Nuclear Physics A}}

% I %%%%%%%%%%%%%%%%%%%%%%%%%%%%%%%%%%%%%%%%%%%%%%%%%%%%%%%%%%%%%%%%%%%%
\section{Introduction}

Current deep inelastic scattering (DIS) experiments at HERA
are probing the structure of the nucleon at ever smaller
values of the Bjorken $x$ variable ($x \equiv Q^2/2M\nu \agt 10^{-4}$),
where $M$ is the nucleon mass, and $Q^2$ and $\nu$ the squared
four-momentum and energy transfer to the nucleon.
The large range of $Q^2$ available ($4 \alt Q^2 \alt 3000$ GeV$^2$)
allows for quantitative tests of the nature of the evolution
of structure functions at these low $x$ \cite{HERA}.
One reason why the small $x$ region is interesting is that
it is here that large distance phenomena become relevant,
which offers the unique prospect of learning more about
the nucleon's non-perturbative structure in DIS.

Before we can fully appreciate the long range structure of
the nucleon, there is of course the non-perturbative region at
moderate $x$, but smaller $Q^2$ ($Q^2 \alt 1$ GeV$^2$),
which has been the focus of many previous studies, but which
still requires quantitative understanding.
The significance of this kinematic region is that it reflects
precisely the transition between the perturbative and
non-perturbative domains of QCD, namely the interface between
the parton picture of the nucleon, and the familiar valence quark
models of hadrons which successfully describe much of the
low-energy phenomenology.

This connection has previously been investigated within
phenomenological models of the nucleon, such as bag or
non-relativistic quark models \cite{MODELS}.
While these efforts have produced some encouraging results,
some problems remain, however, in bridging the gap between
leading-twist quark distributions calculated within such models,
and the experimental structure function data.
One problem encountered in most of these approaches has been
their failure to correctly describe the $x \rightarrow 0$
behavior of valence quark distributions \cite{REGGE}.
This is intrinsically related to the common approximation made
in identifying the non-interacting quark system, which is formally
on-mass-shell and remains spectator to the deep inelastic
collision of the probe with the interacting quark, with a
simple diquark, of mass $m_S \sim (2/3) M$.
A consequence of this assumption is that to ensure correct
normalization, the calculated valence quark distribution must
be evolved from extremely low resolution scales,
$Q^2 \sim$ 0.1--0.3 GeV$^2$ \cite{MODELS}, where
$\alpha_s \sim$ 1--3.
At these scales the use of perturbative QCD is rather
questionable, and it's not even clear whether adding
next-to-leading-order corrections \cite{GRV,FT,NLO}
is a sensible solution \cite{WEST}.
In this paper we will address the question of whether,
and under what conditions, can the inclusion of higher mass
spectator states generate sufficiently soft contributions to
the valence quark densities so as to allow reliable descriptions
of the data when evolved from a scale $Q^2 \sim 1$ GeV$^2$,
where perturbative QCD can be used with some more justification.

Our physical framework is assumed to be as follows.
At a scale $Q^2 \sim 1$ GeV$^2$ it will be natural to view the
nucleon as composed of constituent quark (CQ) ``quasi-particles'',
which for our purposes will mean quarks having the quantum numbers
of QCD (current) quarks, but which acquire a large average mass,
$m_Q \sim M/3$, due to their strong non-perturbative interactions.
Since this scale is also typically that associated with chiral
symmetry breaking, pseudoscalar Goldstone bosons (pions) will in
addition play a role.
We will therefore assume that the structure of the nucleon seen by
a virtual photon with $Q^2 \sim 1$ GeV$^2$ is incorporated within
the substructure of the CQs themselves and their pion dressing.

In addition to knowing the internal structure of CQs, one also
needs to describe the interaction between CQs in the nucleon,
which we parametrize in terms of a relativistic quark--nucleon
vertex function.
We should state that our aim here is not to calculate the vertex
function directly, and it will suffice to adopt a parametrization
based on a physically motivated ansatz.
Our main aim will be to model the spectral function which describes
the non-interacting, spectator quark system, in terms of the degrees
of freedom available to us at $Q^2 \sim 1$ GeV$^2$, namely CQs and
pions.
Within our framework the task of describing the recoil spectator
spectrum of the nucleon is reduced to detailing the remnant recoil
system that remains after the interaction with the CQ.

The outline of this paper is as follows.
In Section \ref{SSS} we introduce our model for the spectral
function of the spectator quark system.
In Section \ref{IA} we discuss the CQ momentum distribution in the
nucleon and the low mass spectator spectrum, which gives the
leading contribution to the nucleon structure function at
large $x$.
The intermediate mass contributions, which are modeled by
dressing the CQs by pions, are described in Section \ref{cloud}.
Also discussed is the small-$x$ region of the quark distributions,
for which we use a simple model based on Regge phenomenology to
describe the large spectator mass continuum contribution to the
spectral function.
Finally, in Section \ref{finita} we discuss our numerical results
and draw conclusions.

% II %%%%%%%%%%%%%%%%%%%%%%%%%%%%%%%%%%%%%%%%%%%%%%%%%%%%%%%%%%%%%%%%%%%
\section{Spectrum of Spectator States}
\label{SSS}

In this Section we outline the basic reasoning behind the need
for a detailed description of the spectrum of states that are
spectator to the interaction of the photon with a quark in the
nucleon (see Fig.1).
Let us begin the discussion with the familiar hadronic tensor
$W_{\mu\nu}$ which describes deep-inelastic scattering of a
virtual photon (momentum $q$) from a nucleon target (momentum $P$):
\begin{eqnarray}
W_{\mu\nu}(P,q)
&=& {1\over 4\pi} \int d^4\xi\ e^{i q\cdot \xi}
    \left< P | J_{\mu}(\xi) J_{\nu}(0) | P \right>,
\end{eqnarray}
where $J_{\mu}$ is the electromagnetic current operator,
and the states are normalized such that
$\langle P \left| \right. P' \rangle
= 2 P_0\ (2\pi)^3 \delta^{4}\left(P-P'\right)$.
For unpolarized scattering, which we consider throughout
this paper, $W_{\mu\nu}$ is usually decomposed into two
independent structures:
\begin{eqnarray}
W_{\mu\nu}(P,q)
&=& \left( -g_{\mu\nu} + { q_{\mu} q_{\nu} \over q^2 }
%     \right) {F_1 \over M}\
    \right) F_1\
 +\ \left( p_{\mu} - q_{\mu} {P\cdot q \over q^2}
    \right)
    \left( p_{\nu} - q_{\nu} {P\cdot q \over q^2}
    \right) {F_2 \over P \cdot q},
\end{eqnarray}
where the structure functions $F_{1,2}$ are expressed
as functions of $x$ and $Q^2 (= - q^2)$.
In the parton model $F_{1,2}$ become independent of $Q^2$
(at leading order in $\alpha_s$), and are related via:
$F_1(x) = F_2(x)/2x = 1/2\ \sum_q e_q^2\ q(x)$,
where $q(x)$ gives the probability to find a quark $q$
in the nucleon.

In the usual treatment of inclusive DIS from a nucleon one
formally sums over the complete set of intermediate states,
which are labeled $n$ in Fig.1, that are spectator to the
deep-inelastic collision.
In order to reliably calculate structure functions for all
$0 < x < 1$ one needs to know the spectator quark spectrum
in its entirety, including the contributions from the large
spectator masses.
In practice this is quite a challenge to any theoretical approach,
and commonly one resorts to approximations which are, however,
often valid only in a limited range of $x$ \cite{MODELS}.

To elucidate the connection between the quark distribution $q(x)$,
calculated at some scale $Q^2 = \mu^2$, and the spectator state $n$,
we can use the so-called dispersion representation of the quark
distribution \cite{LPS,KPW}.
Here $q(x)$ is expressed in terms of the probability density,
$\rho$, of the quark spectator states with invariant mass
$s = (P-p)^2$:
\begin{eqnarray}
\label{qrho}
q(x;\mu^2)
&=& \int ds\int_{-\infty}^{p_{max}^2(s,x)} dp^2\
    \rho(s,p^2,x;\mu^2),
\end{eqnarray}
where $p$ is the four-momentum of the struck quark, and
\begin{equation}
\label{p2max}
p_{max}^2(s,x) = x\left(M^2 - \frac s{1-x}\right)
\end{equation}
is the kinematical maximum of the quark virtuality $p^2$
for given $x$ and $s$.
In terms of the amplitude $\psi_n$ describing the absorption
of a virtual photon by a quark with momentum ${\bf p}$, this
can be written as:
\begin{eqnarray}
\label{rho_psi}
\rho(s,p^2,x;\mu^2)
&=& {1 \over 32\pi^2} {1 \over P\cdot q}
    \sum_n
    \overline{\psi}_n({\bf P}-{\bf p}) \not\!q\
    \psi_n({\bf P}-{\bf p})\ \delta(s - M_n^2),
\end{eqnarray}
where $M_n^2$ is the invariant mass of the intermediate spectator
state $n$, and where formally the amplitude $\psi_n({\bf P}-{\bf p})$
is defined via the matrix elements of the quark field operator at the
origin $\psi(0)$:\
$\psi_n({\bf P}~-~{\bf p})
= \langle {\bf P}~-~{\bf p}, n \left| \psi(0) \right| P \rangle$.
An expression similar to (\ref{qrho}) holds also for the antiquark
distribution $\bar q(x)$.

The dispersion representation in Eq.(\ref{qrho}) is valid as long
as the integrals are convergent.
Therefore it is generally assumed that the spectral density vanishes
at large quark virtualities $p^2$, and that the integration is
dominated by contributions from the region of finite $p^2$,
$|p^2| \alt M^2$ \cite{LPS,KPW}.
Then from Eq.(\ref{p2max}) one sees that the smaller the $x$,
the broader the region of $s$ from which $q(x)$ receives
contributions.
For large $x$ ($x \agt 0.2$), the region of finite $s \alt M^2$ is
of major importance, and here one can expect models with low-mass
spectators, such as diquarks, to be a good approximation.
On the other hand, at small $x$ the quark virtuality $p^2$ is
finite even for large $s \sim  M^2/x$.
Therefore the quark distribution in Eq.(\ref{qrho}) is sensitive
to the high-energy part of the spectrum.

The physical origin of the large-$s$ region of the spectator quark
spectrum is the high-energy scattering of the virtual $q\bar q$
component of the photon from the nucleon.
This can be seen by recalling that the quark field operator
acting on the target state can either annihilate a quark or
create an antiquark in the target.
In the former case the matrix element
$\psi_n\propto\langle {\bf P-p}, n|a({\bf p})|P\rangle$
describes the absorption of a virtual photon by quarks with
momentum ${\bf p}$, which results in a finite spectrum of
intermediate states.
On the other hand, the contribution from the antiquark part
of the field operator,
$\psi_n\propto \langle {\bf P-p}, n|b^{\dagger}(-{\bf p})|P\rangle$,
describes the ``external'' antiquarks from $q\bar q$ fluctuations
of the virtual photon and their interaction with the nucleon.
The possible antiquark momenta ${\bf p}$ are determined by the wave
function of the photon and can be as large as the photon
momentum ${\bf q}$, leading to a large invariant mass of the
antiquark--nucleon system, $s = (P-p)^2 \gg M^2$.

The modeling of the spectral function $\rho(s,p^2,x;\mu^2)$ must
of course reflect the resolution scale at which the target nucleon
is probed.
As mentioned in the Introduction, we will consider $\rho$ at a
fixed scale of order $Q^2 = \mu^2 \sim 1$ GeV$^2$.
At this model scale, constituent quarks are perhaps the most
useful effective degrees of freedom.
In a constituent quark picture, the resolution scale can be
associated with some average size, $R_Q$, of the CQ:
$\mu^2 \sim 1/R_Q^2$.
Results for the color correlation function on a lattice
\cite{LATT} indicate that color fields are confined to a
distance of order 0.2 fm.
This should also apply to each CQ, and one is led to the
estimate $\mu^2 \sim 1$ GeV$^2$.
Naturally, at some other scale, CQs may not be the best variables
with which to describe the spectral function.
However, at $\mu^2 \sim 1$ GeV$^2$, it is known from many other
applications that CQs give a good description of the nucleon's
dynamical properties, such as electromagnetic form factors.
A natural question to ask therefore is whether CQs can be useful
tools in describing DIS from the nucleon, Fig.2.

Within this picture, it will be convenient to break up the
spectrum of all possible spectator states into several mass
regions, which we now summarize.

$\bullet$ (${\cal A}$):\
In a CQ picture, the simplest approximation is to identify the
spectator system which remains after one CQ is removed with a
single diquark, with some relatively low mass, $m_S \sim {\cal O}(M)$,
Fig.3(a).
The low-$s$ part ($s \alt 1$ GeV$^2$) of the spectral function
can then be approximated by:
\begin{eqnarray}
\label{rhoA}
\rho_{\cal A}(s,x,p^2)
&=& \delta\left( s - m_S^2 \right)
f_{Q/N}(x,p^2),
\end{eqnarray}
where the function $f_{Q/N}(x,p^2)$, which is now independent of $s$,
is identified with the distribution of constituent quarks in the
nucleon with a fraction of the nucleon light-cone momentum $x$ and
the invariant mass squared $p^2$ (see Eq.(\ref{fdef}) below)
(we omit explicit reference to the scale $\mu^2 \sim 1$ GeV$^2$
from now on).
In a relativistic treatment, $\rho_{\cal A}$ is best described
in terms of relativistic quark--diquark--nucleon ($QDN$) vertex
functions.
Simple effective models, with constituent quarks as basic
ingredients, can be used to calculate the vertex function
directly.
Note that such calculations are only feasible if the complexity
of the task is reduced to that of a few body problem.
In other models, where the struck quark is treated as a current quark,
the spectator state would have to be very complicated, making
even an approximate model solution for the vertex function
beyond our present technical ability.
Therefore if the spectator state is to be treated as a diquark, then
in order to solve for the relativistic vertex function, the struck
quark must be a constituent quark.

$\bullet$ (${\cal B}$):\
To model the intermediate-mass part (1 GeV$^2$ $\alt s \alt s_0$)
of the spectral function requires going beyond the simplest,
single-spectator diquark approximation for the intermediate state.
One approach would be to consider higher-Fock state components
containing a diquark plus some number of $q\bar q$ pairs
as spectators.
Within the bag model calculations of Ref.\cite{BAG}, for example,
these component were modeled in terms of the insertion of an
antiquark (or quark), produced from the incoming virtual photon,
into the nucleon initial state, which then produced intermediate
states with masses of order $(5/4) M$.

At a scale $Q^2 \sim 1$ GeV$^2$, as well as constituent quarks,
pseudoscalar Goldstone bosons --- pions --- also enter as natural
degrees of freedom in any effective model description.
In a constituent quark picture, therefore, it will be more
convenient to model these components in terms of the dressing
of CQs by pions, as illustrated for example in Fig.3(b).
In practice we restrict ourselves only the one-pion contributions.
One could, in principle, continue the series in the pion number $n$
to large $n$, hoping to describe the large-$s$ region entirely
in terms of pion ladder diagrams.
However, taking this route would necessarily mean eventually including
dressing due to the heavier meson spectators, as well as from diquarks.
A more efficient approach will be to truncate the series
at some mass $s = s_0$ (which we take to be $s_0 \sim 2$ GeV$^2$),
and describe the large-mass contributions in terms of multi-quark
and gluon configurations.

$\bullet$ (${\cal C}$):\
To model the large (including asymptotic) mass tail of intermediate
states ($s \geq s_0 \sim 2$ GeV$^2$), we describe the amplitude for
scattering a $q\bar q$ pair from a CQ within Regge theory, which
effectively takes into account an infinite set of ladders involving
$q\bar q$ pairs in the intermediate state, Fig.3(c).
The exchange of a Regge trajectory with the intercept
$\alpha_R (\approx 1/2)$ leads to imaginary parts of amplitudes
rising with energy as $E^{\alpha_R}$ \cite{REGGE}.
This energy dependence of the scattering amplitude then gives
rise to valence quark distributions which rise at small $x$ as
$x^{-\alpha_R}$.

Within the above framework, all three processes ${\cal A}$--${\cal C}$
will be described in terms of CQ parameters, with the basic common
unit being the relativistic $QDN$ vertex function.
That is, the constituent quark--nucleon interaction is factored
out from the hard scattering of the photon with the interacting
quark.
In the impulse approximation, nucleon DIS can then be viewed as
a two-step process, in terms of photon--CQ, and CQ--nucleon
subprocesses \cite{HWA,ZS,SUZ}.
As we shall see, however, special care has to be given to the
fact that the CQ is bound, and thus off-mass-shell.

% III %%%%%%%%%%%%%%%%%%%%%%%%%%%%%%%%%%%%%%%%%%%%%%%%%%%%%%%%%%%%%%%%%%
\section{Constituent Quark Distributions in the Nucleon}
\label{IA}

At the relatively low momentum scales at which a constituent quark
model may be relevant, higher twists (i.e. $1/Q^2$ corrections to
scaling quark distributions) are bound to play an important role.
The final aim, however, is to compare the calculated structure
functions with DIS data at large $Q^2$ ($Q^2 \agt 5-10$ GeV$^2$),
where higher-twist effects should be negligible.
Therefore it will be sufficient to consider only the leading
twist components of the structure function, represented by the
``handbag'' diagram in Fig.1.
At leading twist, the fields that appear in the formal expressions
for the current commutators in the forward virtual Compton scattering
amplitude are those of point-like, elementary (i.e. current) quarks.
In a constituent quark picture, however, the propagator of the quark
before it is struck is modeled to be that of a quark with mass
$m_Q \sim M/3$ (while the propagating high-momentum quark after
the interaction with the photon is a current quark).
If one attempts to work with effective, composite degrees of freedom
in DIS, a clear and well-defined connection must therefore exist
between current and constituent quarks, in particular with respect
to the nature of the photon--CQ vertex.

An example of a simple dynamical model in which such a link
exists is the Nambu \& Jona-Lasinio (NJL) model \cite{NJL,NJL1}.
Here low-momentum gluons are integrated out and absorbed into
a local, point-like interaction between quarks, characterized
by an effective coupling constant.
(Explicit perturbative gluonic degrees of freedom reappear
when the calculated quark distributions are evolved to the
higher $Q^2$ appropriate for the DIS region.)
The mechanisms used in going from current to constituent quarks
are dynamical quark mass generation, involving the scalar
interaction of the quark with the Dirac vacuum, together
with the spontaneous breaking of chiral symmetry.
The connection between current and constituent quark masses
is formally quantified via a ``gap equation'' \cite{NJL}.
Through interactions with quark condensates, (nearly) massless
current quarks acquire large constituent masses, of order
$m_Q \sim$ 400 MeV.
(For detailed reviews of the NJL approach see Refs.\cite{NJL1}.)

Within this approach a CQ mass therefore appears in the quark
propagator, while one still has a simple $\gamma_\mu$ coupling
at the photon--quark vertex.
There is no form factor suppression at large $Q^2$ arising
from the photon--CQ--current quark vertex.
In a relativistic CQ picture, such as that embodied in the
NJL model, the diagrams responsible for the CQ form factor are
those in which the photon couples to the target CQ via
$q \bar q$ loops.
As $Q^2 \rightarrow \infty$ these rapidly die out, leaving
behind only the scaling part associated with the direct coupling
of the photon to the CQ.
In this way the only difference between scattering from current
and constituent quarks is therefore a renormalization of the
quark mass.
The dressing of the struck quark propagator is not included,
since this would correspond to higher twist corrections.

% III.A ................................................................
\subsection{Quark--Diquark--Nucleon Vertex Functions}

Having made these remarks, we now turn to the calculation
of constituent quark distributions inside the nucleon.
Constituent quarks embedded in a nucleon have some
characteristic momentum distribution, which is essentially
given by the CQ wave function.
Within the present approach it is more conveniently expressed in
terms of the relativistic quark--diquark--nucleon vertex functions,
$\Phi^{(S)}_{QDN}$ (proportional to the product of the wave function
and the quark propagator), connecting the nucleon and off-shell
quark to the non-interacting (spectator) intermediate state diquark.

In the impulse approximation (i.e. no final state interactions
between the struck quark and the residual diquark system) the
CQ distribution (see Eq.(\ref{rhoA})) can be written:
\begin{eqnarray}
\label{fdef}
f_{Q/N}^{(S)}(y,p^2)
&=&
\int {d^4p' \over (2\pi)^4}\,
2\pi \delta\left( y-\frac{p'\cdot q}{P\cdot q} \right)
\delta\left( p'^2 - p^2 \right)
\delta\left( (P-p')^2 - m_S^2 \right)           \nonumber\\
& & \hspace*{2cm} \times
\frac1{2P\cdot q}
{\rm Tr} \left[ \not\!q\ {\cal H}^{(S)}(P,p')
         \right],
\end{eqnarray}
where $y$ is the fraction of light-cone momentum of the nucleon
carried by the CQ, and the operator ${\cal H}^{(S)}$ describes
the soft CQ--nucleon interaction.
The label $(S)$ refers to the possible spin orientations of the
spectator state ($S=0$ or 1), since any realistic model of the
nucleon must incorporate pseudovector as well as scalar diquarks.
In terms of the propagators and $QDN$ vertex functions,
${\cal H}^{(S)}$ is defined as:
\begin{mathletters}
\label{Hdef}
\begin{eqnarray}
{\cal H}^{(S=0)}(P,p)
&=&
{1 \over 2}
(\not\!p - m_Q)^{-1}\ \Phi_{QDN}^{(0)}(P,p)\
(\not\!P + M)\ \overline{\Phi}_{QDN}^{(0)}(P,p)\
(\not\!p - m_Q)^{-1},
\end{eqnarray}
for a scalar ($S=0$) spectator system,
and:
\begin{eqnarray}
{\cal H}^{(S=1)}(P,p)
&=&
{1 \over 2}
(\not\!p - m_Q)^{-1}\ \Phi_{QDN}^{\alpha (1)}(P,p)\
(\not\!P + M)\ \overline{\Phi}_{QDN}^{\beta (1)}(P,p)\
(\not\!p - m_Q)^{-1}\          \nonumber\\
& & \times
\left( -g_{\alpha\beta}
      + { (P-p)_{\alpha} (P-p)_{\beta} \over m_S^2 }
\right),
\end{eqnarray}
\end{mathletters}%
for $S=1$ spectators.
Here $m_Q$ is the CQ mass, and the conjugate vertex is defined as\
$\overline{\Phi}_{QDN}^{(S)}
\equiv \gamma_0\ \Phi_{QDN}^{(S) \dagger}\ \gamma_0$.

The trace in Eq.(\ref{fdef}) can be separated into terms proportional
to the four-momenta $P_{\mu}$ and $p_{\mu}$:
\begin{eqnarray}
\label{f12}
{1 \over 4} {\rm Tr}[{\cal H}^{(S)}(P,p)\ \gamma_\mu]
&=& f^{(S)}_1(p^2)\, P_\mu\
 +\ f^{(S)}_2(p^2)\, p_\mu ,
\end{eqnarray}
where the functions $f^{(S)}_1$ and $f^{(S)}_2$ can be calculated
from the $QDN$ vertex.
In terms of $f^{(S)}_{1,2}$, the constituent quark distribution
function $f^{(S)}_{Q/N}$ is given by:
\begin{eqnarray}
\label{fQN12}
f^{(S)}_{Q/N}(y,p^2)
&=& \frac1{8\pi^2}
\left( f^{(S)}_1(p^2)\ +\ y\ f^{(S)}_2(p^2) \right).
\end{eqnarray}
Note that in general the functions $f^{(S)}_{1,2}$ depend on
$p^2$ as well as $P \cdot p$, however the latter is fixed by the
$\delta$-function in Eq.(\ref{fdef}):
$P \cdot p = (M^2 + p^2 - m_S^2)/2$.

Different $S=0$ and $S=1$ vertices, as well as a larger
pseudovector diquark mass ($m_{S=1}-m_{S=0} \sim 0.2$ GeV),
are needed to explicitly break spin-flavor SU(4) symmetry,
which is reflected in a softer $d$ quark distribution compared
with the $u$ quark distribution \cite{CT}.
The complete Dirac structure for the vertex $\Phi_{QDN}^{(S)}$
can be written down in terms of a number of independent functions.
In particular, one has for the $S=0$ vertex:
\begin{eqnarray}
\label{Phi0}
\Phi_{QDN}^{(0)}(P,p)
&=& \phi^{(0)}\ I\
 +\ \phi_p^{(0)}\ \not\!p,
\end{eqnarray}
and similarly for the $S=1$ case:
\begin{eqnarray}
\label{Phi1}
\Phi_{QDN}^{\alpha (1)}(P,p)
&=& \phi^{(1)}\ \gamma^{\alpha}\gamma_5\
 +\ \phi_p^{(1)}\    p^{\alpha}\gamma_5\
 +\ \phi_P^{(1)}\    P^{\alpha}\gamma_5\
 +\ \cdots ,
\end{eqnarray}
although different bases in which to expand $\Phi_{QDN}^{(S)}$
can of course be chosen \cite{MEYER}.

In most previous calculations of structure functions involving $QDN$
vertices \cite{DM,MSM,MST,MPT,MW} the momentum dependence in
$\Phi_{QDN}^{(S)}$ has not been calculated directly, but rather has
had to be parametrized, by appealing to DIS data to fix parameters.
The approach adopted has been to choose a single form for both the
$S=0$ and $S=1$ vertices, and adjust the momentum dependence in each,
to effectively compensate for the omission of the other structures.
For example, by choosing different shapes for the functions
$\phi^{(S=0)}$, one can get similar results whether one uses a
structure $I$ \cite{DM,MSM,MST} or one proportional to $\not\!p$
(see also Ref.\cite{MW}).

Although the phenomenological approach has been successful
in allowing fair descriptions of the nucleon's valence quark
distributions, it is naturally desirable to have stronger
theoretical justification for using any particular vertex
structure.
Recently attempts have been made to calculate $\Phi_{QDN}^{(S)}$
within effective models, notably the NJL model.
Here we outline the general approaches used in these calculations ---
for more details see Refs.\cite{MEYER,IBY,HT,TUB,SHAKIN}.
The simplest ansatz has been to reduce the relativistic
three-body calculation to a more tractable, effective two-body
problem, by treating the nucleon as a composite of a quark
and an elementary diquark \cite{TUB}.
Within this approximation one can solve exactly the
Bethe-Salpeter equation for the quark--diquark system using
a static approximation for the (local) quark exchange interaction.

Beyond this simplest approximation, the Bethe-Salpeter equation has
more recently been solved for the case where the exchanged quark was
allowed to propagate between the quark and diquark \cite{MEYER,SW}.
Ishii et al. \cite{IBY} and Huang and Tjon \cite{HT} have gone
further by finding solutions to the covariant Faddeev equation
without imposing the constraint that the diquark be elementary.
Although only $S=0$ diquarks were modeled in Refs.\cite{HT,TUB},
the authors of Refs.\cite{IBY,MEYER} have also incorporated
pseudovector diquarks in their models for $\Phi_{QDN}$.
Note that to describe DIS one only needs the vertex function
with an on-shell diquark --- solutions to the Bethe-Salpeter
or Faddeev equations more generally have both the quark
and diquark off their mass shells.

The results of the model calculations \cite{MEYER,SW} indicate
that the $S$-wave vertices give the most important contributions
to the norm and to the nucleon's electric charge, which provides
some justification for the prescription applied in
Refs.\cite{DM,MSM,MST,MPT,MW} of using only the leading
structures.
Furthermore, they show that the shapes of $\phi^{(S)}$
can be reasonably well approximated by simple monopole or
dipole functions:
\begin{eqnarray}
\label{VFpole}
\phi^{(S)}(p^2)\
\propto { \left( m_Q^2 - p^2 \right) \over
          \left( \Lambda_S^2 - p^2 \right)^{n_S} },
\end{eqnarray}
with cut-off parameters $\Lambda_S$ of the
order of 1---2 GeV (depending on the precise value of $n_S$).
Such results are not surprising if one's model is to
reproduce realistic values for the nucleon's static
properties, such as the mass and r.m.s. radius.
Note the presence of the zero in the numerator in
Eq.(\ref{VFpole}) at $p^2 = m_Q^2$, which has the role
of canceling the quark propagators in $f_{Q/N}^{(S)}$.
In principle the quark propagators in Eq.(\ref{fdef})
could develop unphysical poles in the kinematically
allowed region if the sum of quark and diquark masses
falls below the nucleon mass, meaning that the nucleon
could freely disintegrate into its constituents.
As such the form factor in Eq.(\ref{VFpole}) reflects
the underlying dynamics that are responsible for the
bound state of the nucleon.
Similar results have been reported for the
quark--antiquark--meson vertex function in Ref.\cite{CSSSZ}.

Note that $\Lambda_S$ is not related to the intrinsic NJL cut-off,
$\Lambda_{NJL}$.
By treating scalar or pseudovector diquarks as quasi-particles,
with effective boson propagators $\propto ( (P-p)^2 - m_S^2 )^{-1}$,
one renders all loop integrals appearing in the calculation of the
Bethe-Salpeter amplitudes finite, without any need to introduce
ultra-violet cut-offs as regulators.
This can be easily seen by counting inverse powers of momentum
running through the quark--diquark legs.
The cut-off $\Lambda_S$ is therefore only an indicator of the
``hardness'' of the $QDN$ vertex, or the ``size'' of the
quark--diquark configuration in the nucleon.

For the scalar vertex (\ref{Phi0}), the $S$-wave structure\
$\Phi_{QDN}^{(0)} = I\ \phi^{(0)}$ \cite{MSM,MST}
in Eq.(\ref{fdef}) gives for the functions $f^{(0)}_{1,2}$:
\begin{mathletters}%
\label{f12:scalar}
\begin{eqnarray}
f^{(0)}_1(p^2) &=&
{1 \over 2}
{ \left| \phi^{(0)}(p^2) \right|^2 \over (m_Q^2 - p^2) }, \\
f^{(0)}_2(p^2) &=&
{ \left| \phi^{(0)}(p^2) \right|^2 \over \left(m_Q^2 - p^2\right)^2 }\
\left( m_Q M + P\cdot p \right).
\end{eqnarray}
\end{mathletters}%
Note that an equivalent result is obtained from the structure
$\not\!\!P$, or $(I + \not\!\!P/M)/2$, which would be the leading
one in the non-relativistic limit \cite{MEYER}.
With the pseudovector vertex structure:
$\Phi_{QDN}^{(1)} = \gamma_5 \gamma_{\alpha} \phi^{(1)}$
one has:
\begin{mathletters}%
\label{f12:psvector}
\begin{eqnarray}
f^{(1)}_1(p^2) &=&
{1 \over 2}
{ \left| \phi^{(1)}(p^2) \right|^2 \over (p^2-m_Q^2) }
{ (p^2 - M^2 - 2 m_S^2) \over m_S^2 },    \\
f^{(1)}_2(p^2) &=&
{1 \over 2}
{ \left| \phi^{(1)}(p^2) \right|^2 \over (p^2-m_Q^2)^2 }
\left( { (m_Q^2 - M^2) \over m_S^2 } (p^2 - M^2)\
\right.                                         \nonumber\\
& &
\left.
    +\ M^2\ +\ 6 M m_Q\ -\ m_Q^2\ -\ 2 m_S^2\ +\ 2 p^2
\right).
\end{eqnarray}
\end{mathletters}%

{}From Eqs.(\ref{qrho}), (\ref{fdef}) -- (\ref{f12:psvector})
we can now calculate the quark distribution associated
with process ${\cal A}$, which, because of the low mass present
in the spectator state, should be a good approximation
at large values of $x$.

% III.B ................................................................
\subsection{Nucleon Structure Functions at Large $x$}

At large $x$ one can think of the photon scattering
quasi-elastically from the constituent quark, in the
sense that the internal structure of the CQ ``quasi-particle''
is not resolved.
In this case the CQ structure function is a $\delta$-function,
and the resulting nucleon quark distribution given entirely
by the momentum distributions $f_{Q/N}^{(S)}(y,p^2)$.
Integrating $f_{Q/N}^{(S)}$ over $p^2$ gives then the
valence quark distribution in the proton:
\begin{eqnarray}
\label{qA}
q_{\cal A}^{(S)}(x)
&=& \int_x^1 {dy \over y}\
    \int_{-\infty}^{p^2_{max}(m_S^2,y)}\ dp^2\
    f^{(S)}_{Q/N}(y,p^2)\  \delta(1 - x_Q)\
=\  \int_{-\infty}^{p^2_{max}(m_S^2,y)}\ dp^2\
    f^{(S)}_{Q/N}(x,p^2),
\end{eqnarray}
where $x = Q^2/2P\cdot q$,  $x_Q = x/y$ is the Bjorken variable
of the CQ and the maximum value of the quark virtuality is given
by Eq.(\ref{p2max}).

The parameters defining the constituent quark distribution
$f^{(S)}_{Q/N}$ (and hence $q_{\cal A}^{(S)}$) are as follows.
Consistent with typical masses in effective constituent quark
models, we take for the CQ mass $m_Q \approx 450$ MeV.
For the scalar (pseudovector) diquark mass we use
$m_{0(1)} \approx 1.0(1.2)$ GeV.
These diquark masses are somewhat larger than in some model
calculations of the nucleon's static properties, which reflects
the approximation that the low-mass spectator system can be
described as a single diquark, Eq.(\ref{rhoA}).
In fact, one would expect in effective CQ--diquark models to see a
spectrum of diquark bound states with a range of masses \cite{NJL1}.
By taking into account only the lightest diquark state one necessarily
underestimates the average mass of the diquark which would effectively
describe the low-$s$ part of the spectrum.
For comparison, we also examine the effect on the quark distributions
of using smaller masses, $m_{0(1)} \approx 0.6(0.8)$ GeV.

For the $QDN$ form factor (vertex function) specifying the
distributions $f_{Q/N}^{(S=0,1)}$ we use the form in
Eq.(\ref{VFpole}).
The exponential parameters $n_S$ in Eq.(\ref{VFpole}) are fixed
by the behavior of the quark distribution at large $x$, since
\begin{eqnarray}
\label{p2x}
p^2 \rightarrow - {p_T^2 \over (1-x)}\ \ \
{\rm as}\ \ x \rightarrow 1.
\end{eqnarray}
{}From Eqs.(\ref{p2x}) and the expressions for $f_{Q/N}^{(S)}$ one
finds that with $n_0 = 2.0$ and $n_1 = 3.5$ the quark distributions
$q^{(0,1)}(x) \rightarrow (1-x)^{3,4}$ as $x \rightarrow 1$
\cite{PARAM}.
The values of the cut-off parameters $\Lambda_S$ in
Eq.(\ref{VFpole}) are taken to be:\
$\Lambda_0 = 1.0$ GeV for the scalar vertex, and
$\Lambda_1 = 1.2$ GeV for the $\gamma_5 \gamma_{\alpha}$ vertex.

Using these parameters, we show in Fig.4(a) the valence sum
(solid curve),
$x (u_V + d_V) \rightarrow
3 x \left( q_{\cal A}^{(0)} + q_{\cal A}^{(1)} \right) / 2$,
evolved from $\mu^2 = 1$ GeV$^2$ to $Q^2 = 10$ GeV$^2$.
One sees that within the impulse approximation the valence quark
distribution is almost a factor 2 larger than the data \cite{PARAM}
(shaded) at $Q^2 = 10$ GeV$^2$.
The position of the peak value of $x(u_V+d_V)$ depends quite
sensitively on the mass of the intermediate state diquark system.
For the smaller diquark masses ($m_{0(1)} = 0.6 (0.8)$ GeV), the
result (dashed) is nearly a factor 3 larger, and peaks at a
considerably larger value of $x$ compared with the data.
The valence quark ratio in the impulse approximation,
$d_V/u_V \rightarrow
\left. 2 q_{\cal A}^{(1)} \right/
\left( 3 q_{\cal A}^{(1)} + q_{\cal A}^{(0)} \right)$,
shown in Fig.4(b), exhibits far less dependence on the diquark
masses, and gives quite reasonable agreement with the data at
large $x$, particularly for the larger mass spectators.

The fact that the impulse approximation is only good at large $x$
is clearly reflected in the absence of the correct (singular)
behavior of $q_{\cal A}^{(S)}(x)$ as $x \rightarrow 0$ that would
be expected from Regge phenomenology, namely
$\sim x^{-\alpha_R} (\alpha_R \approx 1/2)$ (see Section IV.B).
Because only small (finite) spectator masses are included in
$q_{\cal A}^{(S)}(x)$, the calculated distributions are finite
at $x=0$, whereas the experimental distributions diverge.
Therefore a large fraction of the normalization has to come
from the medium and large-$x$ regions to compensate
--- hence the large peak in $x(u_V+d_V)$.
This fact demonstrates that the distributions $q_{\cal A}^{(S)}(x)$
alone, calculated in a CQ model, are incapable of describing the
valence structure function over the entire range of $x$, and points
to the necessity of including larger spectator masses, corresponding
to processes ${\cal B}$ and ${\cal C}$ outlined in Section II.
In the next Section we describe the contributions to the
structure functions from the larger spectator mass components
of the quark spectral function.

% IV %%%%%%%%%%%%%%%%%%%%%%%%%%%%%%%%%%%%%%%%%%%%%%%%%%%%%%%%%%%%%%%%%%%
\section{Structure of Constituent Quarks}
\label{cloud}

As clearly seen in the previous Section, one should not expect valence
quark distributions calculated from the handbag diagram alone to
exhibit the correct behavior at small $x$.
Since this is intrinsically related to the presence of low-mass
spectators in the intermediate state, to rectify this deficiency
one must consider in addition contributions to the spectral function
$\rho$ from states with masses larger than $m_S \sim M$.

A systematic formulation within the CQ picture developed in Section III
is one in which the interactions among quarks are completely factored
out of the basic $QDN$ vertex.
Interactions between quarks then give rise to effective constituent
quark dressing --- for process ${\cal B}$ by a cloud of pions in the
intermediate-$s$ region, while for process ${\cal C}$ by a soft
``Reggeon cloud'' at large $s$ ---  which would result in CQs attaining
substructure, Fig.2.
The contribution from these processes to the quark distribution
(direct photon diagram, Fig.2(a)) can then be written:
\begin{eqnarray}
\label{qdres}
q^{(S)}_{\cal B,C}(x)
&=& \int {d^4p\over (2\pi)^4}\,
2\pi \delta \left( \left(P-p\right)^2 - m_S^2 \right)
% {\rm Tr}_Q \left[ {\cal Q}(p,q)\
{\rm Tr} \left[ {\cal Q}(p,q)\,
                {\cal H}^{(S)}(P,p)
         \right],
\end{eqnarray}
where the operator
\begin{eqnarray}
\label{Q:def}
{\cal Q}(p,q)
&=& -i \int {d^4k\over (2\pi)^4}\
\delta \left( \left( k+q \right)^2 \right)
% {\rm Tr}_q
\left[
\not\!q\ \left( \not k-m_Q \right)^{-1}\
{\cal T}(p,k)\ \left( \not k-m_Q \right)^{-1}
\right],
\end{eqnarray}
describes DIS from a dressed (generally off-shell) constituent quark.
The trace in Eq.(\ref{qdres}) is taken over the upper and lower quark
indices in Fig.2(a).
The crossed photon diagram in Fig.2(b) gives rise to an antiquark
distribution $\bar q(x)$, which is given by an expression analogous
to Eq.(\ref{Q:def}) with the substitution $q\to -q$.

The operator ${\cal T}(p,k)$ in Eq.(\ref{Q:def}) describes
the truncated quark--quark interaction.
In what follows we assume that ${\cal T}$ is an
analytic function of the invariant mass
squared of the quark--quark system, $s_Q=(p-k)^2$, of $u_Q=(p+k)^2$,
and of the quark virtualities $k^2$ and $p^2$.
For real $s_Q$ and $u_Q$, ${\cal T}$
has a right-hand cut in the variable $s_Q$, a left-hand cut in
$u_Q$, and singularities for $p^2 > 0$ and $k^2 > 0$.
To make use of these properties in the loop integral in
Eq.(\ref{Q:def}), it is convenient to parametrize the loop
momentum $k$ in terms of external momenta $p$ and $q$:
$k=\alpha p + \beta q' + k_T$,
with $q'=x_Q p + q$, $x_Q=-q^2/2p\cdot q$ being the Bjorken
variable of the CQ, and $k_T$ is a two-dimensional vector
perpendicular to both $p$ and $q$.
Then the $\delta$-function in (\ref{Q:def}) fixes $\alpha=x_Q$.
After integrating with respect to $\beta$,
one finds that the operator ${\cal Q}$
vanishes outside the interval $|x_Q|\le 1$.
For $0\le x_Q\le 1$, ${\cal Q}$ is given by a
dispersion integral in $s_Q$
along the right-hand ($R$) cut,
\begin{eqnarray}
\label{Qdisp}
{\cal Q}_R(p,q)
&=& { 1 \over 16\pi^3 } { 1 \over 2 p\cdot q }
\int { d^2k_T\ ds_Q\over 1-x_Q }\,
{
{\rm Im}_R \not\!q (\not\!k + m_Q)
 {\cal T}(p,k) (\not\!k + m_Q)
 \over (k^2 - m_Q^2)^2
},
\end{eqnarray}
and hence contributes to the quark distribution.
For $-1\le x_Q\le 0$, the operator ${\cal Q}$ is given by the
dispersion integral along the (left-hand, $L$) cut in the $u$-channel,
\begin{eqnarray}
\label{QL}
{\cal Q}_L(p,q)
&=& { 1 \over 16\pi^3 } { 1 \over 2 p\cdot q }
\int { d^2k_T\ du_Q\over 1+x_Q }\,
{
{\rm Im}_L \not\!q (\not\!k + m_Q)
 {\cal T}(p,k) (\not\!k + m_Q)
 \over (k^2 - m_Q^2)^2
}.
\end{eqnarray}
${\cal Q}_L$ determines in fact the contribution from the
crossed photon diagram in the physical region $x_Q>0$, and
corresponds to the antiquark distribution in the CQ.
To obtain the antiquark operator $\overline{\cal Q}$ from
Eq.(\ref{QL}) one must substitute $q\to -q$ (and therefore
$x_Q\to -x_Q$) and introduce an overall minus sign,
$\overline{\cal Q}(p,q)=-{\cal Q}_L(p,-q)$ (see e.g. \cite{KPW}).

Before discussing the details of the quark--quark interaction within
our model, we can make one more model-independent observation about
factorization of the total $\gamma^* N$ interaction.
Let us firstly expand the operator ${\cal Q}$ in terms of Dirac
basis tensors:
\begin{eqnarray}
\label{Qdirac}
{\cal Q}(p,q)
&=& I\ {{\cal Q}_0\over 2m_Q}\
 +\ \not\!p\ {{\cal Q}_1\over 2m_Q^2}\
 +\ \not\!q\ {{\cal Q}_2\over 2p\cdot q},
\end{eqnarray}
where the coefficients ${\cal Q}_i$ ($i=0,\ 1,\ 2$)
scalar dimensionless functions of $x_Q$ and $p^2$.
An expression similar to (\ref{Qdirac}) holds for the antiquark
operator with the scalar coefficients $\overline{\cal Q}_{0,1,2}$.
Projecting ${\cal Q}$ onto positive energy space and averaging
over CQ spins one can define the quark and antiquark distributions
in a CQ as:
\begin{mathletters}
\begin{eqnarray}
\label{Qsf}
q_Q(x_Q,p^2)
&=& \frac12{\rm Tr}
\left[ \left(\not\!p + m_Q\right)
       {\cal Q}(p,q)
\right] =
{\cal Q}_0 + {p^2\over m_Q^2}{\cal Q}_1 + {\cal Q}_2,\\
\overline{q}_Q(x_Q,p^2)
&=& \frac12{\rm Tr}
\left[ \left(\not\!p + m_Q\right)
       \overline{\cal Q}(p,q)
\right] =
\overline{\cal Q}_0 +
{p^2\over m_Q^2}\overline{\cal Q}_1 +
\overline{\cal Q}_2.
\end{eqnarray}
\end{mathletters}%
For a quark (antiquark) with no internal structure one can easily
show that in the Bjorken limit only a single function is non-zero:
${\cal Q}_2 \ne 0$ ($\overline{{\cal Q}}_2 \ne 0$),
while ${\cal Q}_{0,1} = 0$ ($\overline{{\cal Q}}_{0,1} = 0$).
On the other hand, for a constituent quark with internal structure
all three coefficients can be non-zero and therefore contribute to
the quark (antiquark) distribution.

Finally, substituting (\ref{Qdirac}) into Eq.(\ref{qdres}) and
using Eqs.(\ref{f12}), (\ref{fQN12}) and (\ref{Qsf}), the contribution
to the nucleon's quark distribution from dressed CQs is:
\begin{eqnarray}
\label{qdres:result}
q^{(S)}_{\cal B,C}(x) &=&
\int_x^1 \frac{dy}y\int^{p_{max}^2(m_S^2,y)}dp^2
\left[
f^{(S)}_{Q/N}(y,p^2) q_Q\left(\frac xy,p^2\right) \right. \\ \nonumber
&&  \left.
+ \Delta f^{(S)}_0(y,p^2) {\cal Q}_0\left(\frac xy,p^2\right)
+ \Delta f^{(S)}_1(y,p^2) {\cal Q}_1\left(\frac xy,p^2\right)
\right],
\end{eqnarray}
where $f^{(S)}_{Q/N}(y,p^2)$ is given by (\ref{fQN12}) while
the functions $\Delta f^{(S)}_{0,1}$ are
\begin{mathletters}
\label{Deltaf}
\begin{eqnarray}
\Delta f^{(S)}_0(y,p^2) &=&
\frac y{32\pi^2m_Q}{\rm Tr}\left[{\cal H}^{(S)}(P,p)\right]
- f^{(S)}_{Q/N}(y,p^2), \\
\Delta f^{(S)}_1(y,p^2) &=& \frac{y\, P\cdot p - p^2}{8\pi^2m_Q^2}
f^{(S)}_1(p^2).
\end{eqnarray}
\end{mathletters}%
A similar expression can be written for the antiquark distribution
$\bar q(x)$ with the substitution ${\cal Q}_i\to \overline{\cal Q}_i$
and $q_Q\to \bar q_Q$.

The first term in Eq.(\ref{qdres:result}) is a generalization of
Eq.(\ref{qA}) for an `undressed' CQ to the case of a CQ having
non-trivial internal structure, and is written as a convolution
of the CQ distribution $f^{(S)}_{Q/N}$ and the spin averaged CQ
structure function $q_Q$, Eq.(\ref{Qsf}).
One sees, however, that the non-trivial Dirac structure of the
$\gamma^* Q$ interaction leads to the convolution breaking terms
in Eq.(\ref{qdres:result}) which thus render the convolution hypothesis
\cite{ZS,BALL,EHQ} invalid.
Note also that a simple convolution formulation of the structure
functions is recovered in the $p^2 \rightarrow m_Q^2$ limit, since
the terms $\Delta f_{0,1}$ come with an extra factor of
$(p^2 - m_Q^2)$ compared with the $f_{Q/N}^{(S)}$ term.

As discussed in Section \ref{SSS}, it is convenient to split the
spectrum of invariant masses of the quark--quark system into two
regions: small $s_Q$ ($s_Q < M^2$) and large $s_Q$ ($s_Q > M^2$).
In what follows we consider a model where the interaction
${\cal T}$ is described by meson exchanges for the
low-mass region, while to evaluate the effect from the large-mass
region we adopt a simple model based on Regge phenomenology.

% IV.A .................................................................
\subsection{Pion-Dressed Constituent Quarks (Process ${\cal B}$)}

In evaluating the effect of the dressing of CQs by meson ``clouds'',
we will restrict ourselves to pions.
Contributions from CQs dressed by higher mass mesons
(e.g. $\rho, \omega$, etc.) will be largely suppressed by the larger
meson masses \cite{MTV}, and only ever become noticeable when
unrealistically hard form factors are employed.
In addition, predictions for the $\rho$ and heavier mesons are less
reliable in constituent quark models due to threshold effects, if one
uses CQ masses of the order $\sim 400$ MeV.

For the interaction of CQs with pions we will utilize the simple
effective chiral Lagrangian, valid at scales below $\sim$ 1 GeV,
similar to that considered in Refs.\cite{MG,EHQ}, in which the
leading interaction is given by:
\begin{eqnarray}
\label{Leff}
{\cal L}_{Q\pi}
&=& - { g_A \over \sqrt{2} f_\pi }\
\overline\psi_Q\ \gamma^\alpha \gamma_5\ \psi_Q\
\partial_\alpha \phi_\pi.
\end{eqnarray}
Here $\psi_Q$ and $\phi_\pi$ are the CQ and pion fields,
$g_A \approx 0.75$ is the axial coupling constant of the
constituent quark, and $f_\pi = 93$ MeV is the pion decay constant.
A typical process which arises from this interaction Lagrangian is
depicted in Fig.5.
Note that the pion fields appearing in ${\cal L}_{Q\pi}$ represent
point-like pions.
In principle ${\cal L}_{Q\pi}$ gives rise therefore also to terms
involving the direct coupling of photons to pions.
At high energies, however, the pion's substructure would be
expected to play a role, and these contributions, through the
presence of the pion form factor, would be suppressed.
Furthermore, the direct $\gamma^* \pi$ couplings would also violate
the Callan-Gross relation, and as such they are not considered.

In the present work we consider only order ${\cal O} (1/f_\pi)$
effects from ${\cal L}_{Q\pi}$.
The relevant process to be considered is therefore the one-pion
exchange process in Fig.6 (together with the crossed photon diagram).
Note that scattering from the pion cloud itself, as would be
needed to generate antiquark distributions, for example, would
enter only at order ${\cal O}(1/f_\pi)^2$ from the Lagrangian
\cite{WW}.
The pion cloud contributions were considered in Ref.\cite{EHQ},
however, the expansion of ${\cal L}_{Q\pi}$ was performed only
up to ${\cal O}(1/f_\pi)$.

The quark--quark operator ${\cal T}$ for the one pion
exchange interaction can then be obtained from (\ref{Leff}) by
a Fierz transformation:
\begin{eqnarray}
\label{Tpi}
{\rm Im}_R {\cal T}^{\pi}
&=& 2\pi \delta \left( (p-k)^2 - m_{\pi}^2 \right)\
{\cal F} \left( \Gamma^{\pi}\ \overline{\Gamma}^{\pi} \right),
\end{eqnarray}
where
$\Gamma^{\pi}
= ( g_A / \sqrt{2} f_{\pi} )
  ( {\not\!p} - {\not\!k} ) \gamma_5 \phi_{Q\pi}$
represents the $QQ\pi$ vertex, with $\phi_{Q\pi}$ being the
$QQ\pi$ vertex function, and ${\cal F}$ is the Fierz transformation
operator.
Substituting (\ref{Tpi}) into (\ref{Qdisp}), we find the one pion
exchange contribution to the coefficient functions ${\cal Q}_i$:
\begin{mathletters}
\label{Qpi}
\begin{eqnarray}
{\cal Q}_0^{\pi}
&=& { 1 \over 8\pi^2 }
\left( {g_A m_Q \over \sqrt2 f_\pi} \right)^2
\int_{-\infty}^{k_{max}^2} dk^2\
{ \phi_{Q\pi}^2 \over (k^2 - m_Q^2)^2 }
\left\{ - x_Q m_\pi^2 \right\},                    \\
{\cal Q}_1^{\pi}
&=& { 1 \over 8\pi^2 }
\left( {g_A m_Q \over \sqrt2 f_\pi } \right)^2
\int_{-\infty}^{k_{max}^2)} dk^2\
{ \phi_{Q\pi}^2 \over (k^2 - m_Q^2)^2 }         \nonumber \\
& & \times
\left\{ (1-x_Q) \left( m_Q^2 - k^2 + x_Q (p^2 - m_Q^2) \right)
- x_Q m_{\pi}^2
\right\},                                        \\
{\cal Q}_2^{\pi}
&=& { 1 \over 8\pi^2 }
\left( {g_A m_Q \over \sqrt2 f_\pi }\right)^2
\int_{-\infty}^{k_{max}^2} dk^2\
{ \phi_{Q\pi}^2 \over (k^2 - m_Q^2)^2 }
\frac 1{m_Q^2}                                  \nonumber \\
& & \times
\left\{
\left( p^2 (2 x_Q - 1) - k^2 \right)
\left( m_Q^2 - k^2 + x_Q (p^2 - m_Q^2) \right)
+ x_Q m_{\pi}^2 \left( p^2 - m_Q^2 \right)
\right\}.
\end{eqnarray}
\end{mathletters}%
The kinematical maximum of the quark virtuality,
$k_{max}^2(s_Q,x_Q,p^2)$, is given by Eq.(\ref{p2max}),
with $s_Q=m_{\pi}^2$ and the nucleon mass squared $M^2$
replaced by $p^2$.

Equations (\ref{Qpi}) determine the pion contribution to the quark
distribution in the CQ via Eq.(\ref{qdres:result}).
Similar expressions arise for the pion contribution to the antiquark
distribution, $\overline{\cal Q}_i^\pi$, and are related to (\ref{Qpi})
by crossing symmetry, viz. interchanging $x_Q\to -x_Q$ and introducing
an overall minus sign.
In the present work we consider the pion contribution to the
valence part of the nucleon quark distribution, so that only
terms even in $x_Q$ survive.
In this case the scalar term ${\cal Q}_0^{\pi}$ cancels out
and the convolution breaking term in Eq.(\ref{qdres:result})
arises only from the Dirac structure ${\not\!p}$ in
Eq.(\ref{Qdirac}).

In Eqs.(\ref{Qpi}) the integral over $k^2$ is regularized by a sharp
cut-off, $\Lambda_{Q\pi}$, whereby the $QQ\pi$ vertex function is
parametrized by a theta-function:
\begin{eqnarray}
\label{VFpiQ}
\phi_{Q\pi}(k^2,p^2)
&=& \Theta \left( k^2 + \Lambda_{Q\pi}^2 \right)\
    \Theta \left( p^2 + \Lambda_{Q\pi}^2 \right),
\end{eqnarray}
where we have also parametrized the $p^2$ dependence of this
vertex by a similar cut-off (for other forms of the cut-off
see Refs.\cite{EHQ,SST,SHAKPI}).
The parameter $\Lambda_{Q\pi}$ represents a typical scale beyond
which the effective interaction in Eq.(\ref{Leff}) is no longer
a valid approximation to QCD, and is usually taken to be of the
order $\Lambda_{Q\pi} \sim 1$ GeV \cite{MG}.

In Fig.7 we plot the one-pion exchange contribution to the
sum of the valence
$u_V \left( \rightarrow
       3/2\ \left( q_{\cal B}^{(0)} - \bar q_{\cal B}^{(0)} \right)
     + 5/2\ \left( q_{\cal B}^{(1)} - \bar q_{\cal B}^{(0)} \right)
     \right)$
and
$d_V \left( \rightarrow
         3\ \left( q_{\cal B}^{(0)} - \bar q_{\cal B}^{(0)} \right)
       + 2\ \left( q_{\cal B}^{(1)} - \bar q_{\cal B}^{(1)} \right)
     \right)$
distributions, compared with the impulse approximation contribution,
for three values of the ultra-violet cut-off $\Lambda_{Q\pi}$,
namely 0.8, 1.0 and 1.2 GeV.
The results indicate that for reasonable values of $\Lambda_{Q\pi}$
($\alt 1$ GeV), the one-pion exchange contribution is certainly not
negligible for $x \alt 0.2$, and indeed provides around 15-20\% of
the valence quark normalization compared to the impulse approximation
distributions associated with process ${\cal A}$.
(Similar numbers are obtained in non-covariant, infinite momentum
frame calculations, where one makes use of a transverse momentum
cut-off,
$\phi_{Q\pi}
= \Theta \left( k_T^2 - \Lambda_{Q\pi}^2 \right)$
--- see Appendix A.)
However, it is also apparent that DIS from dressed CQ is not
able to provide sufficiently soft contributions to the nucleon
quark distribution that would simulate Regge behavior.
Agreement with large $Q^2$ DIS data would still require evolution
from scales similar to that needed for process ${\cal A}$ alone,
$\mu^2 \sim 0.2$ GeV$^2$ \cite{MW}.
Adding higher order contributions would make the total distribution
softer, however, one would need to include quite high orders to
come close to reproducing the required degree of singularity as
$x \rightarrow 0$.
The complexity involved in performing even the 2-loop calculation
\cite{WW} based on the effective interaction ${\cal L}_{Q\pi}$ is
quite daunting!
In the next Section we shall examine a more efficient method
of incorporating the larger mass continuum of spectator states
(above $s \sim 2$ GeV$^2$), based on Regge phenomenology.

% IV.B .................................................................
\subsection{Asymptotic Spectator Masses (Process ${\cal C}$)%
%            from High Energy Quark--Quark Scattering (Process ${\cal C}$)%
\protect\label{regge}}

It should be clear by now that to obtain quark distributions which
are sufficiently soft at small $x$ to be compatible with DIS data
(when evolved from our model scale of $\mu^2 \sim 1$ GeV$^2$) one
must include contributions to the spectator spectral function from
the large-mass continuum.
As mentioned earlier (see also Ref.\cite{KPW}), the physical
mechanism to which one can attribute the large mass states is
the high-energy scattering of the $q\bar q$ components of the
photon from the nucleon.
By examining the structure of the energy denominators of
the $q\bar q + N$ system in the target rest frame one can
immediately see that this time-ordering dominates the contribution
from the direct scattering of the photon from a quark in the target.
The virtual $q\bar q$ fluctuation excites the target into states
with a spectrum of masses $s \agt {\cal O}(M^2)$.

In modeling the large-$s$ contribution to the spectral function
it will be convenient to make use of well-known ideas from the Regge
poles approach to high-energy scattering \cite{REGGE}.
Formally, the sum over the spectator masses $s$ can be shown to
be effectively described by the exchange of Regge trajectories
in the complex angular momentum plane.
In a constituent quark model, where the interaction of the
photon with the CQ is factored from the total amplitude, the
$q\bar q$ pair will in fact excite the CQ in the nucleon,
which necessitates integrating over states that are spectator
to the $\gamma^* Q$ collision.
The problem then reduces to describing the asymptotic behavior of
the quark--quark interaction ${\cal T}(p,k)$
in terms of exchange of Regge poles, Fig.3(c)
(for an alternative description of high-energy quark--quark
scattering see Ref.\cite{GW}.)

The leading contribution to the asymptotic scattering amplitude
comes from the Pomeron exchange, which, however, cancels out in
the valence quark distribution.
The relevant Regge trajectories therefore correspond to the
exchange of vector mesons, and we will assume that the
corresponding Reggeons couple to quarks like vector mesons.
In this case the quark--quark scattering amplitude at high energies
can be written:
\begin{eqnarray}
\label{T:def}
{\cal T}(p,k) &=&
\gamma_{(q)\mu}\ \gamma_{(Q)}^\mu
{ T(s_Q,u_Q;k^2,p^2) \over s_Q - u_Q },
\end{eqnarray}
where $T$ is a Lorentz-invariant function, whose interpretation
becomes clear if we project ${\cal T}$ onto positive
energy space and average over quark spins:
\begin{eqnarray}
T(s_Q,u_Q;k^2,p^2)
&=& \frac14 {\rm Tr}_q{\rm Tr}_Q
\left[
\left( - \not\!k + m_Q \right) \left( \not\!p + m_Q \right)
{\cal T}(p,k)
\right].
\end{eqnarray}
Here the trace is taken over the upper and lower quark indices
in Fig.2, which we distinguish by indices $(q)$ and $(Q)$,
respectively.
Therefore the function $T(s_Q,u_Q;k^2,p^2)$ can be considered
as an analytical continuation of the spin-averaged quark--quark
scattering amplitude into the off-mass-shell region.
In general $T$ describes both the $q Q$ and $\bar q Q$ scattering
channels, which are connected by crossing symmetry.
Namely, in the $s$-channel $T(s_Q,u_Q;k^2,p^2)$ coincides with the
$\bar q Q$ scattering amplitude, $T_{\bar qQ}(s_Q;k^2,p^2)$, which
contributes to the quark distribution, while in the $u$-channel it is
identified with the $q Q$ amplitude, $T_{qQ}(u_Q;k^2,p^2)$, from which
one obtains the antiquark distribution.

For the interaction (\ref{T:def}), the coefficients
${\cal Q}_{0,1,2}^R$ in the high-energy (Regge) region become:
\begin{mathletters}
\label{QR}
\begin{eqnarray}
{\cal Q}_0^R
&=& 0,                                                          \\
{\cal Q}_1^R
&=& {1 \over (2\pi)^3}
% \int ds_Q\int_{-\infty}^{k^2_{max}(s_Q,x_Q,p^2)} dk^2\,
\int ds_Q\int_{-\infty}^{k^2_{max}} dk^2\,
{ {\rm Im}T_{\bar qQ}(s_Q;k^2,p^2) \over \left( k^2 - m_Q^2 \right)^2 }
\left\{ { 2m_Q^2 x_Q^2 \over s_Q - p^2 - k^2 } \right\},        \\
{\cal Q}_2^R
&=& {1 \over (2\pi)^3}
% \int ds_Q\int_{-\infty}^{k^2_{max}(s_Q,x_Q,p^2)} dk^2\,
\int ds_Q\int_{-\infty}^{k^2_{max}} dk^2\,
{ {\rm Im}T_{\bar qQ}(s_Q;k^2,p^2) \over \left( k^2 - m_Q^2 \right)^2 }
\left\{ -x_Q + {m_Q^2 - k^2 - 2x_Q^2p^2\over s_Q - p^2 - k^2} \right\},
\end{eqnarray}
\end{mathletters}%
where $k_{max}^2 = k^2_{max}(s_Q,x_Q,p^2)$ is given by Eq.(\ref{p2max}),
with the nucleon mass squared $M^2$ replaced by $p^2$.
Together with Eq.(\ref{qdres:result}), these give the contribution
to the nucleon quark distribution from the $q\bar q$ scattering
mechanism.
Expressions for the antiquark distributions,
$\overline{\cal Q}_{0,1,2}^R$, can be obtained by applying the
crossing symmetry rules.
The final result is similar to Eqs.(\ref{QR}), with the
$\bar qQ$ scattering amplitude $T_{\bar qQ}$ replaced by the
$qQ$ scattering amplitude $T_{qQ}$.

To finally calculate the Regge-exchange contribution to the quark
($q_{\cal C}$) and antiquark ($\bar q_{\cal C}$) distributions from
Eqs.(\ref{qdres:result},\ref{QR}) requires modeling the $\bar qQ$
and $qQ$ scattering amplitudes.
As discussed in Ref.\cite{KPW}, at high energy and low $x$ one can
approximate $T_{\bar qQ}$ and $T_{qQ}$ by the constituent quark--target
amplitudes obtained from the simple additive quark model of high-energy
hadron scattering.
In this case, the nucleon--nucleon scattering amplitude can be
expressed in terms of the quark--quark amplitude and the CQ
distribution functions $f_{Q/N}$:
\begin{eqnarray}
\label{TQ}
{ T_{NN'}(s_N) \over s_N }
&=& \int [dp]\,[dp']\
f_{Q/N}\left(y,p^2\right)\ f_{Q/N}\left(y',p^{'2}\right)\
{ T_{QQ'}\left(s_Q,p^2,p^{'2}\right) \over s_Q }
\end{eqnarray}
where $s_N$ is now the total hadronic center of mass energy squared,
$p$ and $p'$ are the four-momenta of CQs in nucleons $N$ and $N'$,
with $y$ and $y'$ the light-cone momentum fractions carried by quarks
$Q$ and $Q'$, and $s_Q=(p+p')^2$ is the invariant mass squared of
the interacting $QQ'$ system.
The brackets in Eq.(\ref{TQ}) denote:
$\int [dp] \equiv \int_0^1 dy/y \int_{-\infty}^{p_{max}(m_S^2,y)}dp^2$.
At high energy, where $s_Q \approx y\ y'\ s_N$,
one can obtain approximate solutions to
Eq.(\ref{TQ}) by evaluating the quark--quark amplitude at averaged
values of $y$ and $p^2$ \cite{KPW}.
With the $QDN$ vertex model parameters in Section III,
one finds $\langle y \rangle \sim 1/3$ and
$\langle p^2 \rangle \sim -(0.3-0.4)$ GeV$^2$.

The $s_Q$-dependence of the amplitudes corresponding to
the valence quark distribution, namely the difference
$\Delta T_{QQ} \equiv T_{\bar qQ} - T_{qQ}$, is then fixed
directly by the energy dependence of the $\bar pp$ and $pp$
amplitude difference.
In the Regge approach this difference is determined by the
exchange of the spin-1 $\omega$-meson trajectory:
\begin{eqnarray}
\label{DTpp}
{\rm Im}\left( T_{\bar pp}(s_N) - T_{pp}(s_N)
        \right)
&=& s_N^{\alpha_\omega}\ R_\omega,
\end{eqnarray}
where $R_{\omega}$ is the residue of the $\omega$ Reggeon trajectory,
and $\alpha_\omega$ the intercept.
The best fit of Ref.\cite{DL} gives
$R_{\omega}\approx 42\,{\rm mb}\cdot{\rm GeV}^{1-\alpha_\omega}$
and
$\alpha_\omega=0.548$ .
For the $p^2$ and $k^2$ dependence of $\Delta T_{QQ}$ we follow
Ref.\cite{KPW} in adopting the following factorized form:
\begin{eqnarray}
\Delta T_{QQ}(s_Q;k^2,p^2)
&=& g_R(k^2)\ g_R(p^2)\
\Delta T_{QQ}(s_Q;0,0),
\end{eqnarray}
where the form factor functions $g_R$ describe the fall-off
with four-momentum of the total quark--quark amplitude:
\begin{eqnarray}
g_R(p^2)
&=& \left( 1 - p^2/\Lambda_R^2 \right)^{-n_R}.
\end{eqnarray}
For the exponent $n_R$ we choose the value $n_R \approx 4$ to
correctly reproduce the tail of the resulting distribution
function at large $x$, while the cut-off $\Lambda_R$ remains
a free parameter, to be determined by fitting to the DIS data.
The best fit to the data gives $\Lambda_R \approx 0.25$ GeV.

Finally, with these ingredients the contributions to the $u_V$ and
$d_V$ valence quark distributions from process ${\cal C}$ are both
given by
$3/2 \left(
     \left( q_{\cal C}^{(0)} - \bar q_{\cal C}^{(0)} \right)
   + \left( q_{\cal C}^{(1)} - \bar q_{\cal C}^{(1)} \right)
     \right)$,
where $q_{\cal C}^{(0,1)}$ are obtained from
Eqs.(\ref{qdres:result}), (\ref{Deltaf}) and and (\ref{QR}).

% V %%%%%%%%%%%%%%%%%%%%%%%%%%%%%%%%%%%%%%%%%%%%%%%%%%%%%%%%%%%%%%%%%%%%
\section{Results and Conclusions}
\label{finita}

With the inclusion of all three processes ${\cal A, B}$ and ${\cal C}$,
describing the low, intermediate and high-mass spectrum of spectator
states, respectively, the valence $u$ and $d$ quark distribution are
given by:
\begin{mathletters}
\begin{eqnarray}
u_V(x)
&=& Z_u
    \left\{ u_{\cal A}(x)\
        +\ \left( u_{\cal B}(x) - \overline u_{\cal B}(x) \right)\
        +\ \left( u_{\cal C}(x) - \overline u_{\cal C}(x) \right)
    \right\},                    \\
d_V(x)
&=& Z_d
    \left\{ d_{\cal A}(x)\
        +\ \left( d_{\cal B}(x) - \overline d_{\cal B}(x) \right)\
        +\ \left( d_{\cal C}(x) - \overline d_{\cal C}(x) \right)
    \right\}.
\end{eqnarray}
\end{mathletters}%
The normalization constants $Z_{u,d}$ are determined by
charge conservation,
\begin{mathletters}
\begin{eqnarray}
Z_u &=& { 2 \over
          2 + \langle u-\bar u \rangle_{\cal B}
            + \langle u-\bar u \rangle_{\cal C} }, \\
Z_d &=& { 1 \over
          1 + \langle d-\bar d \rangle_{\cal B}
            + \langle d-\bar d \rangle_{\cal C} },
\end{eqnarray}
\end{mathletters}%
where the brackets $\langle \cdots \rangle$ denote the first moment,
and where the impulse approximation contributions, $d_{\cal A}$ and
$u_{\cal A}$ are normalized to 1 and 2, respectively.

The relative normalizations of the individual contributions from
${\cal A} - {\cal C}$ are completely fixed in terms of the parameters
described in Sections III and IV.
The contributions to the normalization of the valence $u_V+d_V$
distribution from one pion exchange is $\approx 16\%$, while the
Regge exchange contribution is $\approx 28\%$.
This means that the impulse approximation contribution (process
${\cal A}$) is renormalized by $Z_u = 62\%$ and $Z_d = 43\%$ for
the $u$ and $d$ distributions, respectively, and contributes
about 56\% to the total normalization.
The contributions to the second moments of $u_V+d_V$ at the input
scale are $\approx$ 41\%, 6\% and $0.06$\%, so that some 53\% of
the nucleon's momentum at $\mu^2 = 1$ GeV$^2$ is left to be carried
by sea quarks and gluons.

The results for $x(u_V + d_V)$, evolved from $\mu^2 = 1$ GeV$^2$,
are plotted in Fig.8, in comparison with the data (shaded) at
$Q^2 = 10$ GeV$^2$.
Also shown for comparison are the corresponding results for
processes ${\cal A}$ and ${\cal B}$, and ${\cal A}$ alone.
In total, the results indicate that one can indeed obtain a good
description of data above $\mu^2 = 1$ GeV$^2$ in terms of CQ
parameters.
This scale, being at the boundary of the scales which are
generally accepted as reliable for perturbative evolution,
is considerably larger than those used in many previous
attempts to calculate leading twist quark distributions
(which typically have needed $\mu^2 \sim 0.1-0.3$ GeV$^2$).
This is one of the main results of the present work.

As possible extensions of this work, one could next attempt to
incorporate sea quark distributions into the formalism at small
$x$, which would require a more detailed analysis of processes
${\cal B}$ and ${\cal C}$.
In the former, the sea would simply be associated with the
antiquark content of the pion cloud.
However, to adequately describe the low-$x$ region one would
need to extend the Reggeon model for process ${\cal C}$ to
include in addition the exchange of Pomeron trajectories between
quarks.
Another possible test of the model would be to consider polarization
observables, such as the nucleon's spin-dependent $g_1$ structure
function.

Finally, we should state that while our formalism is quite general,
it is nevertheless a formidable challenge to embed within any
specific model the full spectator mass spectrum, so as to describe
consistently the nucleon's structure function over the whole range
of Bjorken-$x$.
Our attempt here to obtain such a unified description suggests,
however, that constituent quarks may still be the most relevant
degrees of freedom appropriate for this task, and for the more
general problem of understanding the nucleon's non-perturbative
structure in DIS.

%%%%%%%%%%%%%%%%%%%%%%%%%%%%%%%%%%%%%%%%%%%%%%%%%%%%%%%%%%%%%%%%%%%%%%%%
\acknowledgements

We would like to thank H.Forkel, G.Piller, C.M.Shakin and C.Stangl
for useful comments and discussions.
This work was supported by the BMBF.

\appendix
%%%%%%%%%%%%%%%%%%%%%%%%%%%%%%%%%%%%%%%%%%%%%%%%%%%%%%%%%%%%%%%%%%%%%%%%
\section{Constituent Quarks in the Infinite Momentum Frame}

In this Appendix we discuss an alternative, non-covariant,
formulation of DIS from constituent quarks to that in Section IV.
In a covariant formalism contributions to the quark distribution
functions from DIS off dressed CQs can be expressed as a
two-dimensional convolution (in $y$ and $p^2$) of the CQ
distribution $f_{Q/N}$ and the off-shell CQ structure function
$q_Q$, plus a non-convolution term whose contribution is proportional
to $(p^2 - m_Q^2)$.
In the on-mass-shell limit this term would vanish, leaving only
the convolution term.
Furthermore, the remaining term would reduce to a one-dimensional
convolution in the momentum fraction $y$ only.
{}From a practical point of view, this would considerably reduce the
number of integrations needed, and simplify calculations involving
higher-order, multi-loop, pion-dressing contributions.

A framework in which one-dimensional convolution equations can
be recovered is non-covariant time-ordered perturbation theory
in the infinite momentum frame (IMF).
Because here all particles are on-mass-shell, the non-convolution
off-shell corrections in Eqs.(\ref{qdres:result}) are identically
zero, and factorization of subprocesses is automatic.
Furthermore, in the IMF the so-called $Z$-diagrams, which involve
antiparticles, or particles moving backwards in time, are suppressed
by powers of $1/P_L$ \cite{WEIN}, where $P_L$ is the longitudinal
momentum of the target.
Hence only forward moving particles are ever considered.
It is only within this approach that one can consider the function
$f_{Q/N}$ (integrated over transverse momentum) as a genuine
probability distribution function --- in any other frame one must
also incorporate antiquark components into the virtual constituent
quark (the IMF is also the most appropriate context in which to
view the $q \rightarrow q \pi$ splitting functions of Eichten et al.
\cite{EHQ,BALL}, and the nucleon $\rightarrow$ nucleon + $\pi$
distributions of Refs.\cite{DLY,MTV}.)
Therefore the simple one-dimensional convolution formulation is
necessarily frame-dependent.

By way of illustration, let us examine the one-pion exchange
contribution to the quark distribution function, as in Fig.6,
in the IMF.
The momentum of a nucleon moving with longitudinal momentum
$P_L$ in the IMF can be parametrized as \cite{WEIN,DLY}:
\begin{mathletters}
\label{Ppar}
\begin{eqnarray}
P_{\mu}
&=& \left( P_L + {M^2 \over 2 P_L}; {\bf 0}_T, P_L \right).
\end{eqnarray}
Similarly for the CQ and spectator diquark momenta, we have:
\begin{eqnarray}
\label{ppar}
p_{\mu}
&=& \left( |y| P_L + {m_Q^2 + p_T^2 \over 2 |y| P_L};\
           {\bf p}_T,\ y P_L
    \right),    \\
p'_{\mu}
&=& \left( |1-y| P_L + {m_S^2 + p_T^2 \over 2 |1-y| P_L};\
           {\bf -p}_T,\ (1-y)\ P_L
    \right),
\end{eqnarray}
and for the struck quark and spectator pion:
\begin{eqnarray}
k_{\mu}
&=& \left( |x| P_L
           + {m_Q^2 + \left( {\bf k}_T + {x\over y} {\bf p}_T \right)^2
           \over 2 |x| P_L};\
           {\bf k}_T + {x \over y} {\bf p}_T,\ x P_L
    \right),    \\
k'_{\mu}
&=& \left( |y-x| P_L
           + {m_{\pi}^2
             + \left( -{\bf k}_T
                      + \left( 1-{x\over y} \right) {\bf p}_T
               \right)^2 \over 2 |y-x| P_L};\
    \right.                                     \nonumber\\
& & \left. \hspace*{2cm}
           -{\bf k}_T + \left( 1-{x\over y} \right) {\bf p}_T,\
           (y-x) P_L
    \right),
\end{eqnarray}
\end{mathletters}%
respectively.
Evaluating the trace as in Eq.(\ref{Q:def}) in terms of these momenta,
one finds the contribution to the valence quark distribution from
a constituent quark dressed by a pion:
\begin{eqnarray}
q_{\cal B}^{(S)}(x)
&=& \int_x^1 {dy \over y}\
    \int_0^{\infty} dp_T^2\ f_{Q/N}^{(S)}(y,p_T)\
    \int_0^{\infty} dk_T^2\ q_{Q\pi/Q}(x_Q,k_T),
\label{qpi}
\end{eqnarray}
where in the IMF the functions $f_{Q/N}^{(S)}$ are given by:
\begin{eqnarray}
f_{Q/N}^{(0)}(y,p_T)
&=& { 1 \over 16 \pi^2 }
    { \left| \phi^{(0)}(y,p_T) \right|^2
      \over y^2\ (M^2 - M^2_{QD})^2 }
    \left\{ p_T^2\ +\ (m_Q + y M)^2 \right\}
\label{fQN0imf}
\end{eqnarray}
for $S=0$ spectators, and
\begin{eqnarray}
f_{Q/N}^{(1)}(y,p_T)
= { 1 \over 16 \pi^2 }
    { \left| \phi^{(1)}(y,p_T) \right|^2
      \over y\ (M^2 - M^2_{QD})^2 }
    \left\{ 6 M m_Q
          + 2 P \cdot p
          + {4 P \cdot p'\ p\cdot p' \over m_S^2}
    \right\}
\label{fQN1imf}
\end{eqnarray}
for $S=1$ intermediate states.
The variable
\begin{eqnarray}
M^2_{QD} &\equiv& (p + p')^2\
        =\ { m_Q^2 + p_T^2 \over y }\
        +\ { m_S^2 + p_T^2 \over 1-y }
\label{sQD}
\end{eqnarray}
is the squared mass of the virtual quark--spectator
diquark system.
Note that the IMF expressions for $f_{Q/N}^{(S)}(y,p_T)$
can be related to the covariant expressions in Section III
(apart from the $QDN$ vertex function)
if one observes that $p_\mu$ in Eq.(\ref{ppar}) satisfies:
\begin{eqnarray}
p_\mu p^\mu &=& - { p_T^2 \over 1-y }\ +\ p^2_{max},
\label{p2}
\end{eqnarray}
where $p^2_{max}$ is given by Eq.(\ref{p2max}).

The function:
\begin{eqnarray}
\label{qQpi}
q_{Q\pi/Q}(x_Q,k_T)
&=& { 1 \over 8 \pi^2 }
    \left( { g_A m_Q \over \sqrt{2} f_{\pi} } \right)^2
    {\left| \phi_{Q\pi}(x_Q,k_T) \right|^2
     \over x_Q^2 (1-x_Q) (m_Q^2 - M^2_{Q\pi})^2}
    \left\{ k_T^2 + m_Q^2 (1-x_Q)^2 \right\}
\end{eqnarray}
represents the (unintegrated) structure function of a CQ
dressed by a pion (see Eqs.(\ref{Qsf}), (\ref{Qpi})),
where the squared mass of the $Q \pi$ state is given by:
\begin{eqnarray}
M^2_{Q\pi}
\equiv (k+k')^2\
 =\ {m_Q^2 + k_T^2 \over x_Q}\
 +\ {m_{\pi}^2 + k_T^2 \over 1-x_Q}.
\end{eqnarray}
One can easily demonstrate that a pseudoscalar $Q\pi$ interaction
leads to the same result as in Eq.(\ref{qQpi}), by virtue of the
IMF on-mass-shell condition for the CQs (i.e. $p_\mu p^\mu = m_Q^2$
in Eq.(\ref{Qpi})), and the Goldberger-Treiman relation \cite{NJL1}.

Although the kinematical (or trace) factors are similar in the
covariant and IMF calculations, the connection between the vertex
functions (or form factors) in the two approaches is not a
straightforward one.
Covariantly, the vertex function such as in Eq.(\ref{VFpole})
can only depend on the virtuality of the exchanged
off-mass-shell particle, $p^2$.
In the time-ordered IMF approach the vertex function cannot
depend on the $p^2$, but rather must be a function of the
variable $M_{QD}^2$.
This constraint stems from the requirement that the
momentum distribution functions respect probability
and momentum conservation, hence must display explicit
symmetry under the interchange $y \leftrightarrow 1-y$
(Eq.(\ref{fDy}) in Appendix B below), which cannot
be the case in a covariant formulation.

Since the $p_T$ and $k_T$ dependence in Eq.(\ref{qpi}) is
factorized, one can write the total quark distribution as a
one-dimensional convolution of the CQ structure function, $q_Q$,
with the $p_T$-integrated CQ distribution $f_{Q/N}^{(S)}$
(c.f. Eq.(\ref{qdres:result})):
\begin{eqnarray}
\label{qBcon}
q_{\cal B}^{(S)}(x)
&=& \int_x^1 { dy \over y }\
    \widetilde{f}_{Q/N}^{(S)}(y)\ q_Q\left({x\over y}\right),
\end{eqnarray}
where
\begin{eqnarray}
\widetilde{f}_{Q/N}^{(S)}(y)
&=& \int dp_T^2\ f_{Q/N}^{(S)}(y,p_T),
\end{eqnarray}
and
\begin{eqnarray}
q_Q(x_Q)
&=& \int dk_T^2\ q_{Q\pi/Q}(x_Q,k_T).
\end{eqnarray}

The convolution formula (\ref{qBcon}) can also be easily generalized
to the case of $n$ pions in the intermediate state, as in Fig.3(b):
\begin{eqnarray}
q_{(n)}^{\pi (S)}
&=& \int {dy_1 \over y_1} \cdots {dy_n \over y_n}
    \widetilde{f}_{Q/N}^{(S)}(y_1)\
    q_Q(y_2/y_1)\ \cdots q_Q(y_n/y_{n-1})\ q_Q(x/y_n).
\end{eqnarray}
As mentioned above, a formulation of the higher-order diagrams in
terms of simple convolution integrals enables fewer integrations
over coupled momenta to be performed, which simplifies considerably
the numerical evaluation of these contributions.

%%%%%%%%%%%%%%%%%%%%%%%%%%%%%%%%%%%%%%%%%%%%%%%%%%%%%%%%%%%%%%%%%%%%%%%%
\section{Diquark Structure}

In this Appendix we discuss DIS from possible diquark constituents
in the nucleon.
If one takes seriously the quark--diquark picture of the
nucleon, in which the diquark itself is treated as an effective
quasi-particle, then one must also consider the process where
a diquark is struck by the DIS probe, Fig.9, which then requires
modeling the structure function of a bound diquark.
(Although any realistic model of the nucleon, which is to reproduce
the Callan-Gross relation, cannot have diquarks as elementary
constituents.)
For scalar diquarks this process was considered by Suzuki et al.
\cite{SUZ}.
In a covariant framework scattering from $S=1$ diquarks is more
problematic, however, as was demonstrated in Ref.\cite{MTV}
for DIS off spin-1 meson configurations in the nucleon.
Within the IMF approach discussed in Appendix A above, calculating
the diquark distribution is considerably simpler --- in fact one
expects explicit symmetry between the constituent quark and diquark
distribution functions:
\begin{eqnarray}
\label{fDy}
f_{D/N}^{(S)}(y) &=& f_{Q/N}^{(S)}(1-y),
\end{eqnarray}
where $f_{D/N}^{(S)}(y)$ is the probability distribution of
spin $S$ diquarks in the nucleon with momentum fraction $y$.
In a covariant treatment this symmetry, which is `screened'
by off-mass-shell effects, need not be explicit.

The relation in Eq.(\ref{fDy}) is readily obtained from any
$QDN$ vertex function that is a function of the variable
$M^2_{QD} \equiv (m_Q^2 + p_T^2)/y\ +\ (m_S^2 + p_T^2)/(1-y)$
(see Eq.(\ref{sQD})) ---
for example a multipole form:\
$\phi^{(S)} \propto 1 / (\Lambda_S^2 + M^2_{QD})^{n_S}$.
With the interchange of $m_Q \leftrightarrow m_S$ and
$y \leftrightarrow 1-y$, this choice automatically results
in the relation (\ref{fDy}) \cite{MW,MTV,ZOL}.
This constraint on the CQ--diquark symmetry consequently
imposes constraints upon the shape of the quark distribution.
Phenomenologically, the main difference between these forms
arises at small $x$, where the $M^2_{QD}$-dependent functions
give somewhat smaller distributions.
The reason for this is the $1/y$ factor in $M^2_{QD}$, which
at small $x$ serves to suppress the quark distributions,
which themselves depend on inverse powers of $M^2_{QD}$.

Having outlined how one could systematically include DIS
contributions from diquarks, we should point out however that
DIS from diquarks is in fact a next-to-leading order process
in the $QQ$ coupling constant $G$.
Hence it should be treated on the same footing as the order
$(1/f_\pi)^2$ processes involving pion dressing, such as the
direct scattering from the nucleon's pion cloud.
While the higher-order (higher-mass) processes can in principle
be systematically included, it is in fact more efficient to
describe them in terms of the Regge model described in Section IV.B.

%%%%%%%%%%%%%%%%%%%%%%%%%%%%%%%%%%%%%%%%%%%%%%%%%%%%%%%%%%%%%%%%%%%%%%%%

% \newpage

\begin{figure}
\centering{\ \epsfig{figure=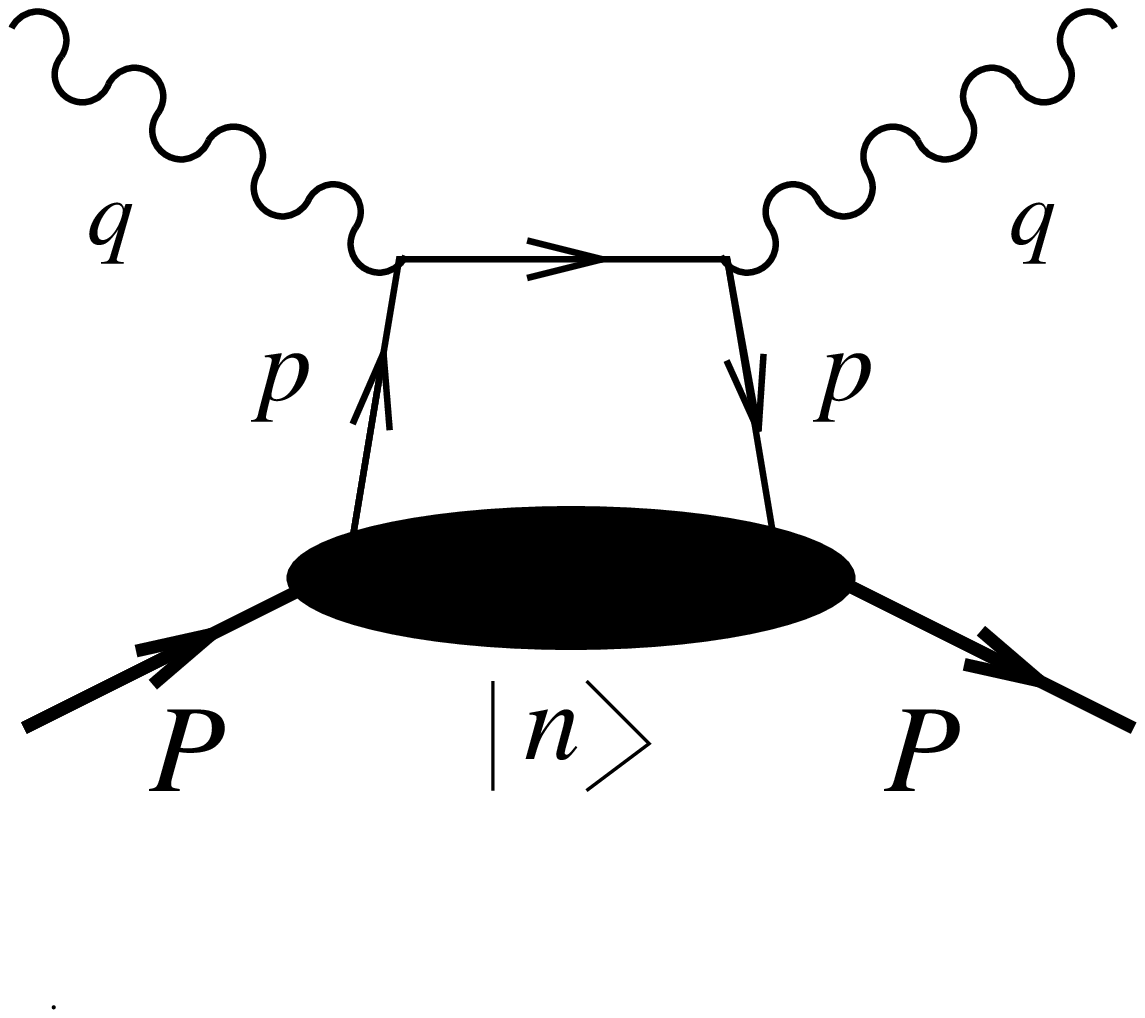,height=10cm}}
\caption{Leading twist contribution to deep-inelastic scattering
         of a photon (momentum $q$) from a constituent quark ($p$)
         within a nucleon ($P$), with $| n \rangle$ labeling the spectator
         quark system.}
\label{F1}
\end{figure}

\newpage

\begin{figure}
\centering{\ \epsfig{figure=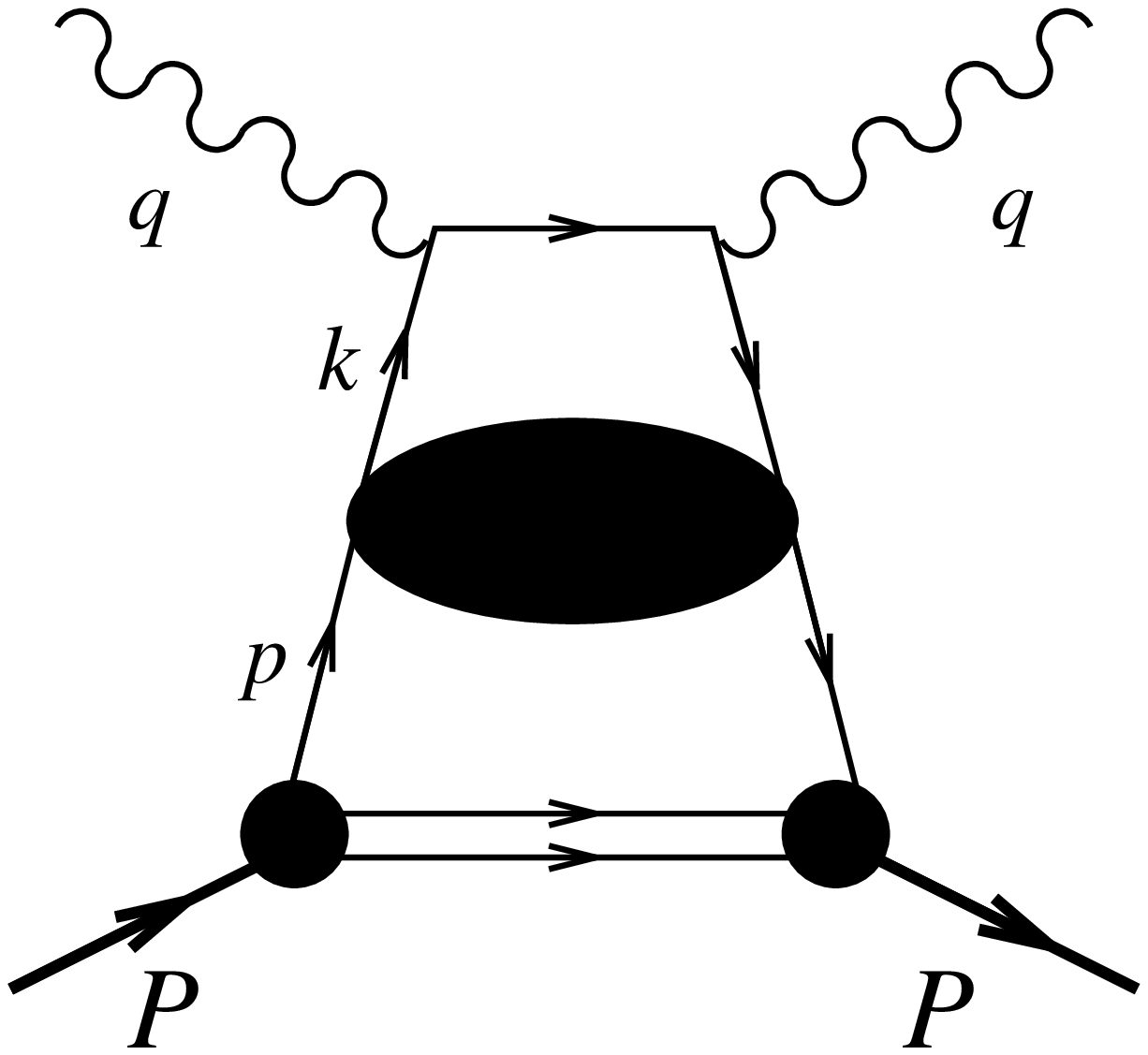,height=9cm}}
\centering{\ \epsfig{figure=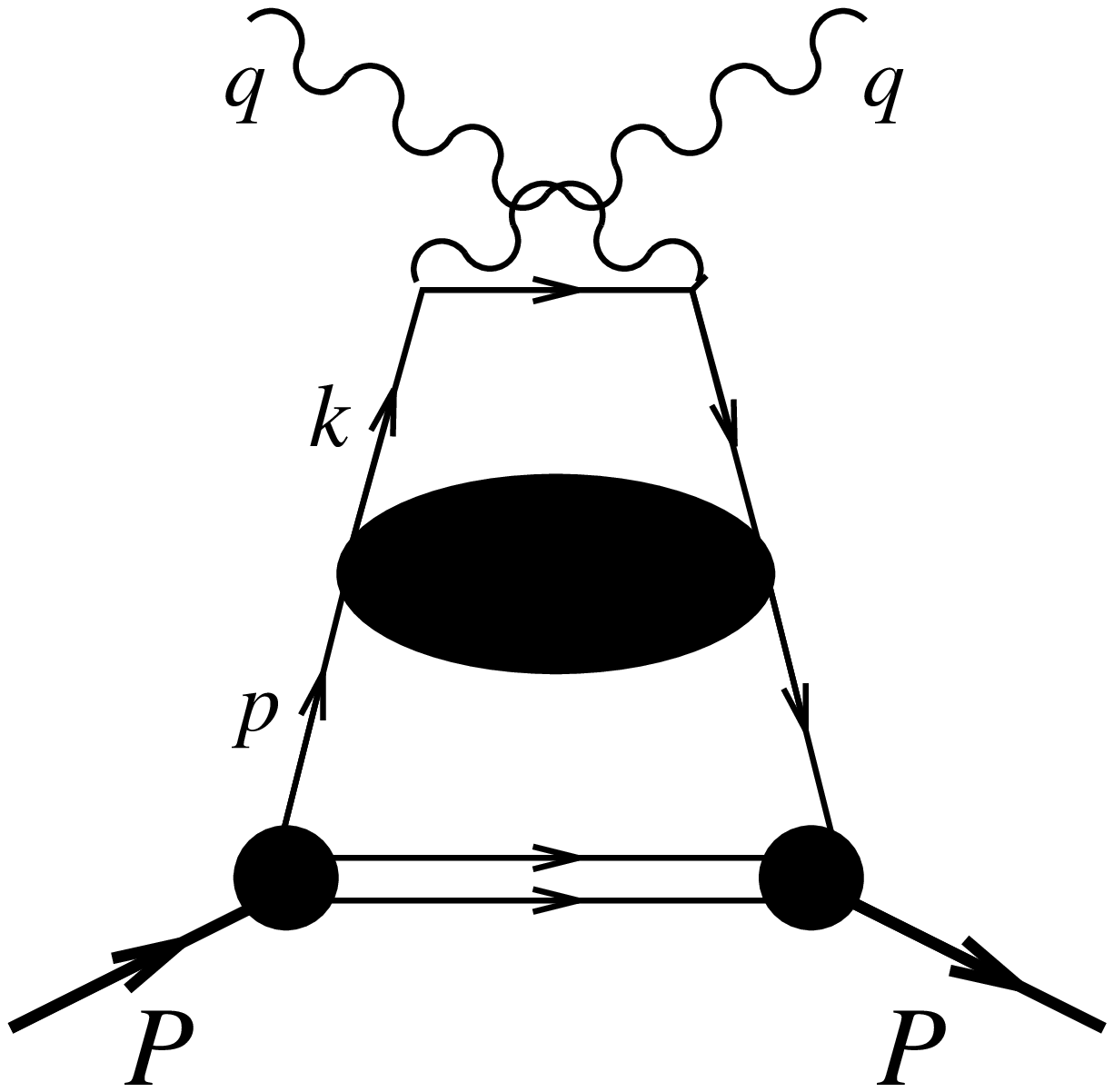,height=9cm}}
\caption{DIS from a constituent quark with substructure.
         (a) direct photon diagram, which contributes to the
         quark distribution,
         (b) crossed photon diagram, associated with the antiquark
         distribution in the nucleon.}
\label{F2}
\end{figure}

\newpage

\begin{figure}
\centering{\ \epsfig{figure=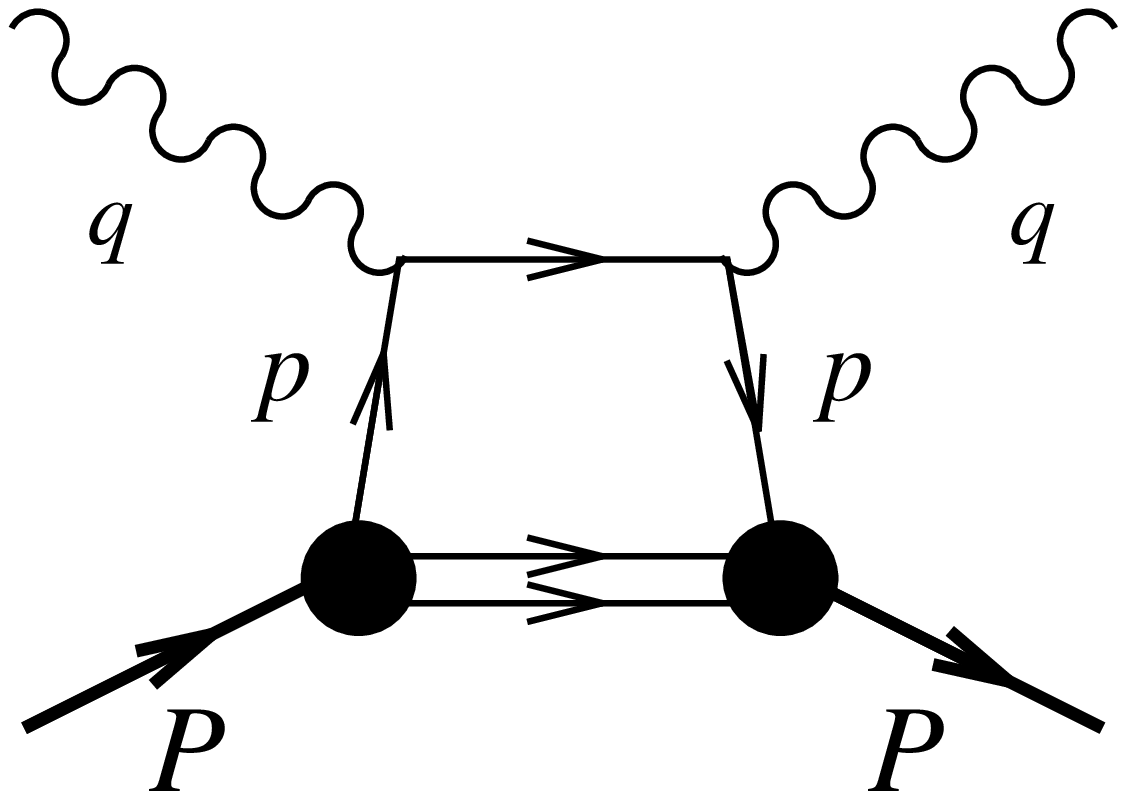,height=8cm}}
\centering{\ \epsfig{figure=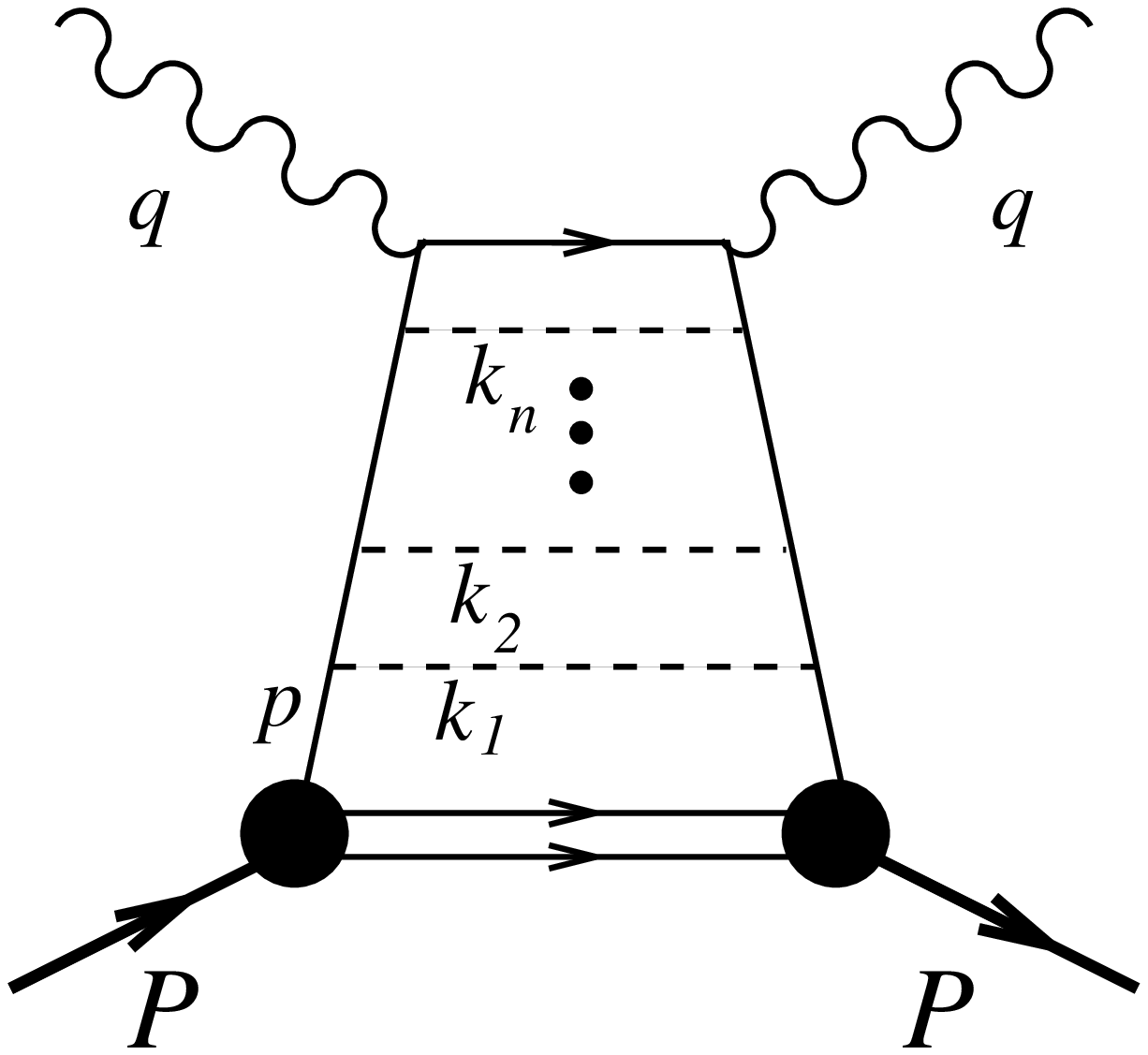,height=8cm}}
\centering{\ \epsfig{figure=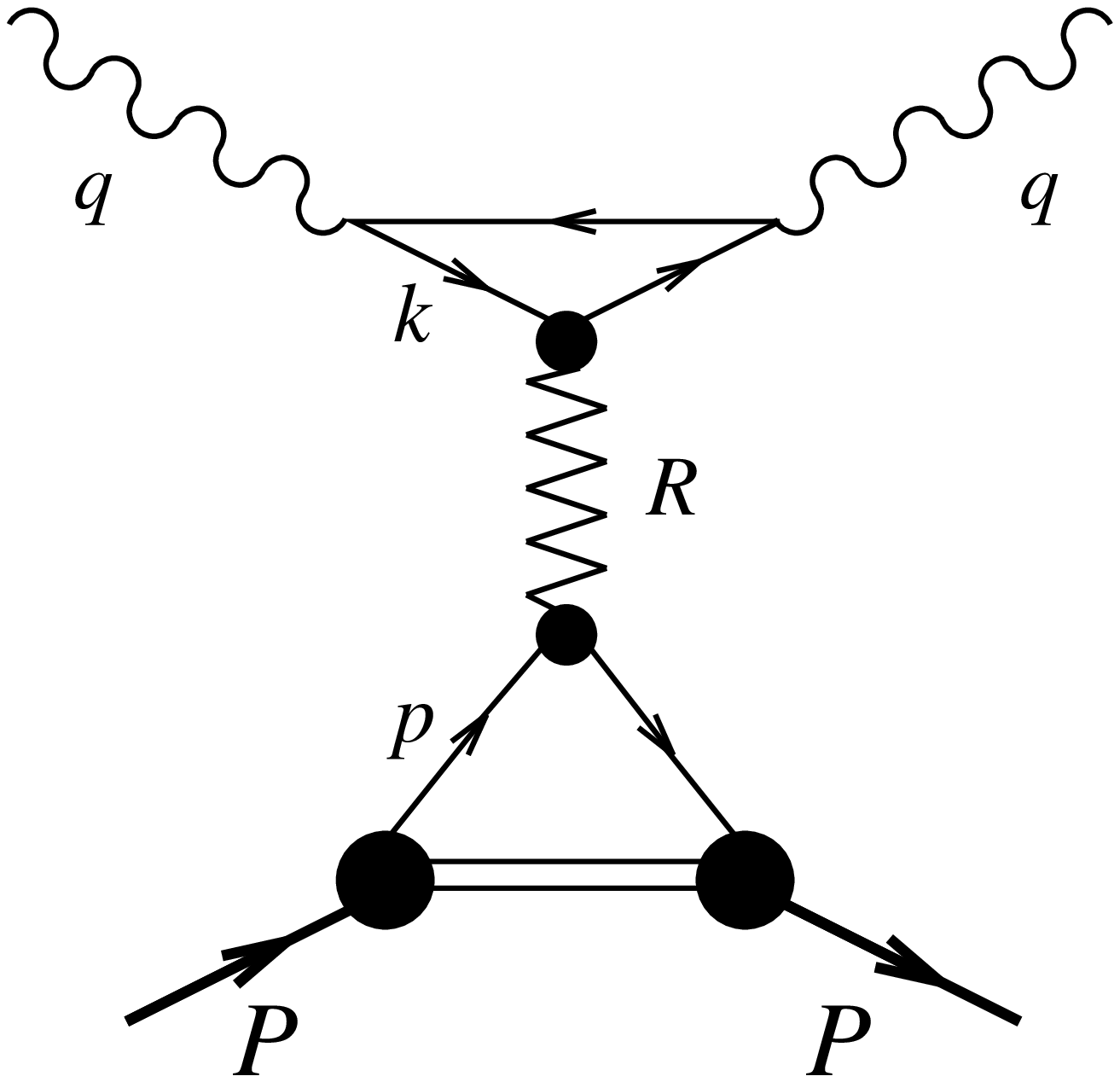,height=8cm}}
\caption{Contribution to the nucleon structure function from the
         (a) low mass spectator (diquark),
         (b) intermediate mass pion-exchange, and
         (c) large mass (asymptotic)
         components of the nucleon spectral function.}
\label{F3}
\end{figure}

\newpage

\begin{figure}
\centering{\ \epsfig{figure=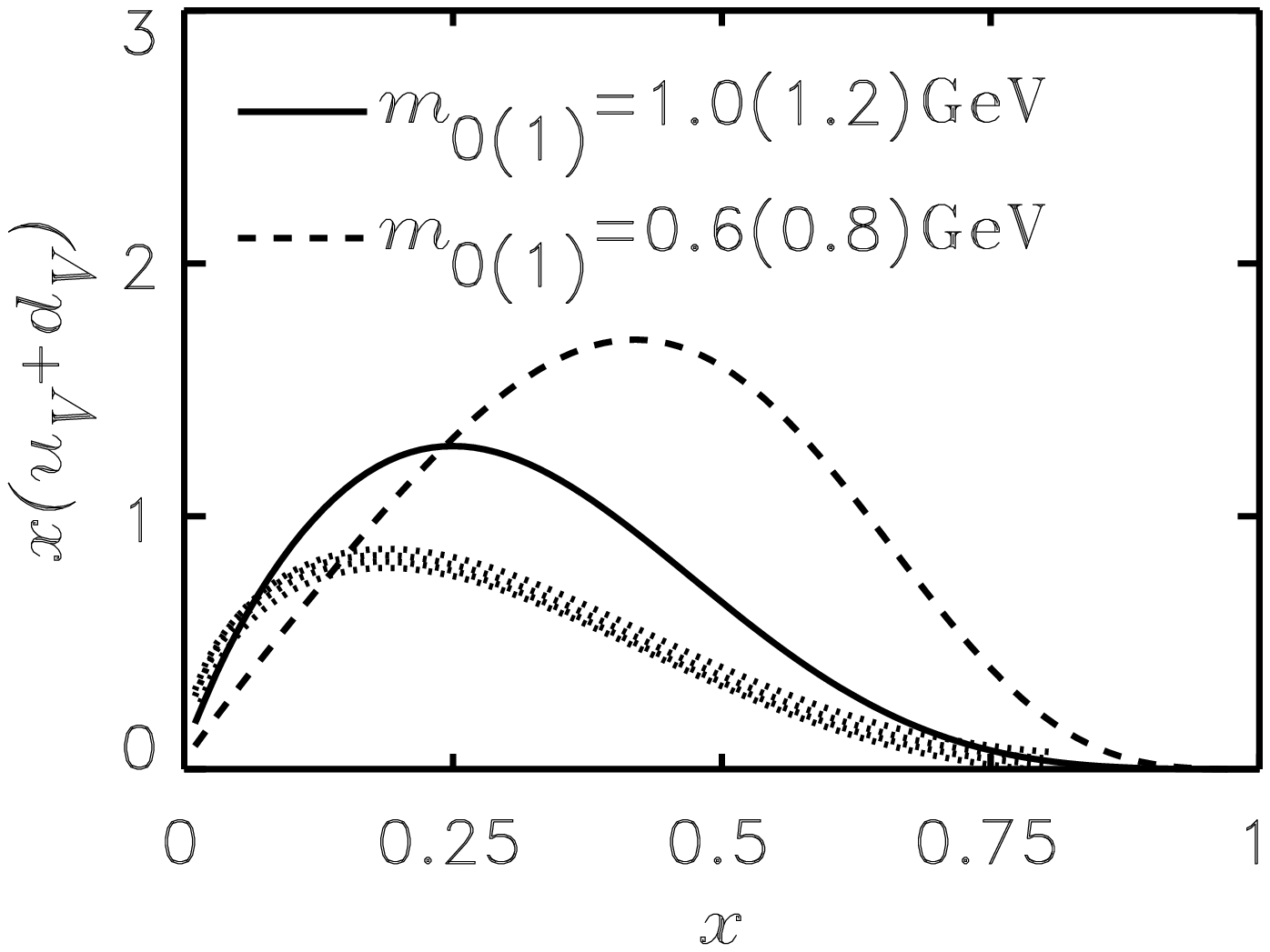,height=9cm}}
\centering{\ \epsfig{figure=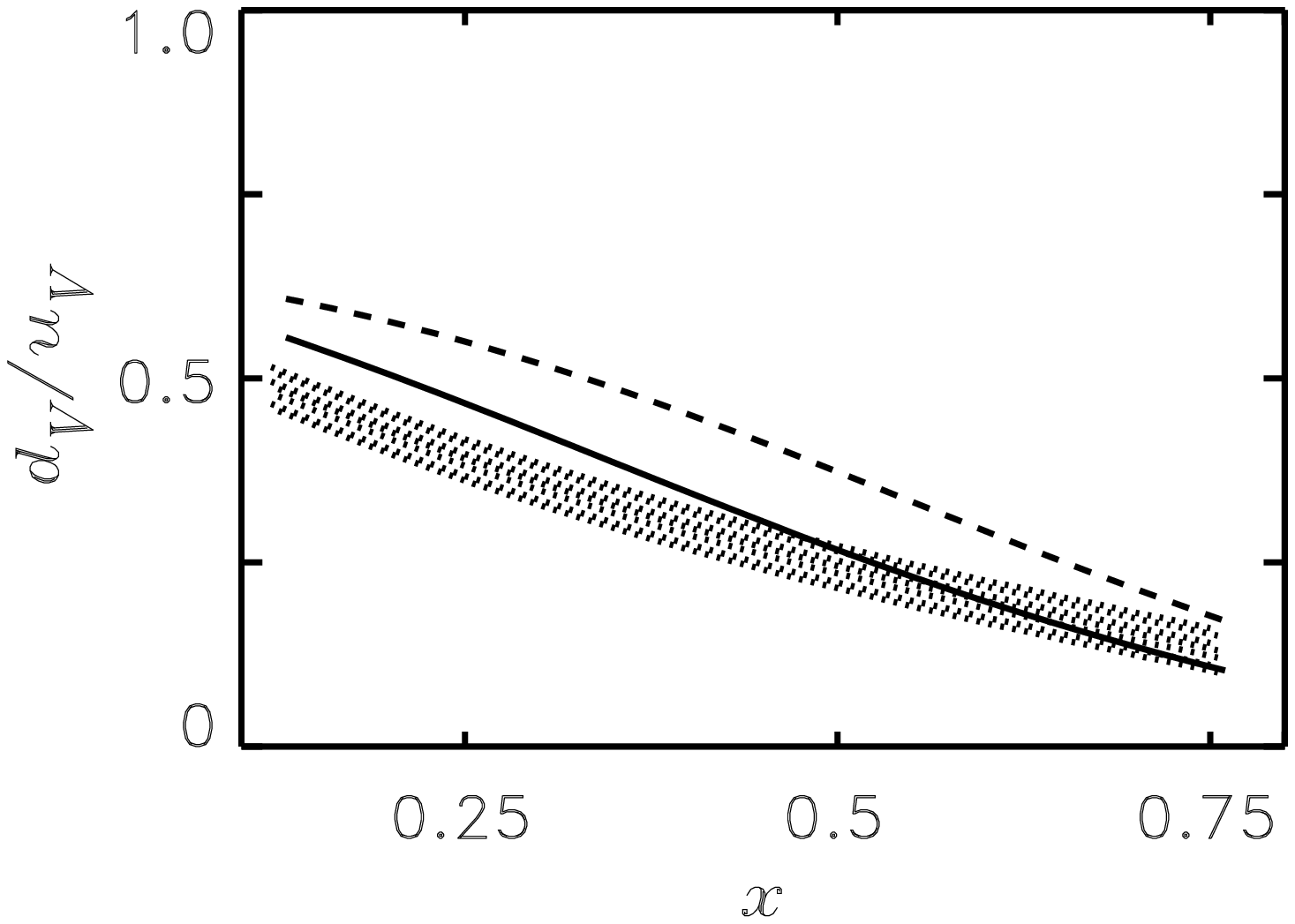,height=9cm}}
\caption{(a) Total valence $x(u_V + d_V)$ quark distribution,
         and (b) valence $d_V/u_V$ ratio,
         including only diquark spectators (process ${\cal A}$),
         with two sets of mass parameters,
         $m_{0(1)} = 1.0 (1.2)$ GeV (solid) and
         $m_{0(1)} = 0.6 (0.8)$ GeV (dashed).
         The curves are evolved from $\mu^2 = 1$ GeV$^2$ to
         $Q^2 = 10$ GeV$^2$, and the shaded area represents
         parametrizations of data taken from
         Ref.\protect\cite{PARAM}.}
\label{F4}
\end{figure}

\newpage

\begin{figure}
\centering{\ \epsfig{figure=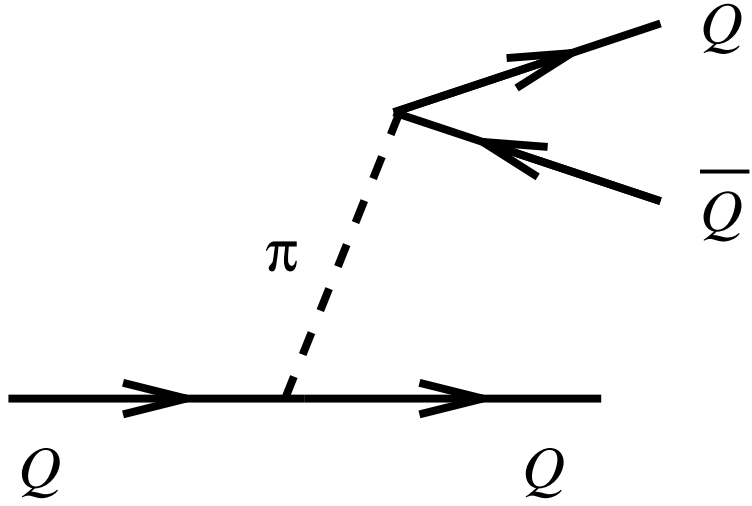,height=8cm}}
\caption{Typical dissociation process generated from the effective
         constituent quark--pion interaction ${\cal L}_{Q\pi}$.}
\label{F5}
\end{figure}

\newpage

\begin{figure}
\centering{\ \epsfig{figure=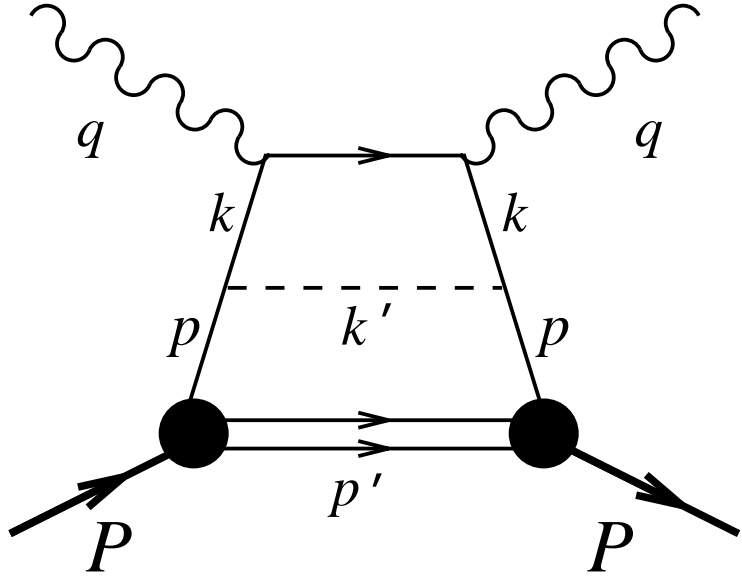,height=9cm}}
\caption{One-pion exchange contribution to the quark distribution
         function.}
\label{F6}
\end{figure}

\newpage

\begin{figure}
\centering{\ \epsfig{figure=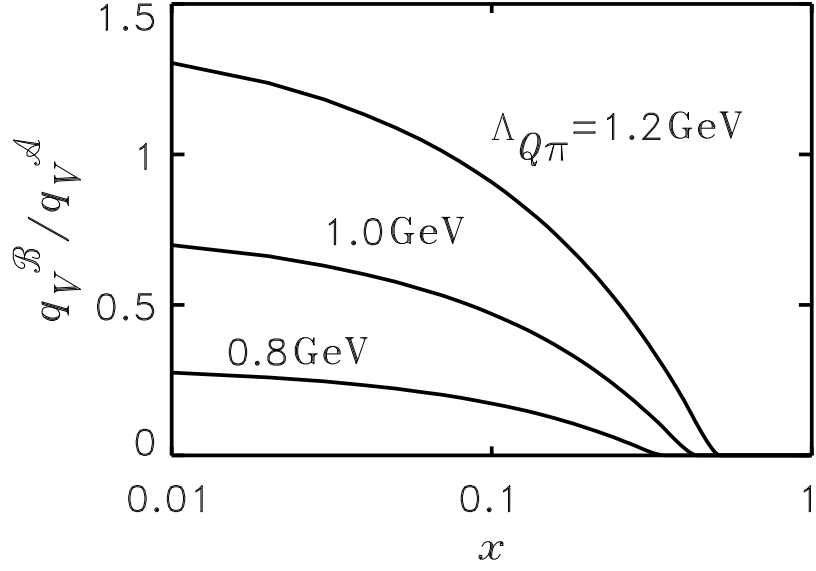,height=9cm}}
\caption{Ratio of the one-pion exchange (${\cal B}$) to the handbag
         contributions (${\cal A}$) to the total valence distribution,
         $q = u_V + d_V$, for different $Q\pi$ vertex cut-off parameters.}
\label{F7}
\end{figure}

\newpage

\begin{figure}
\centering{\ \epsfig{figure=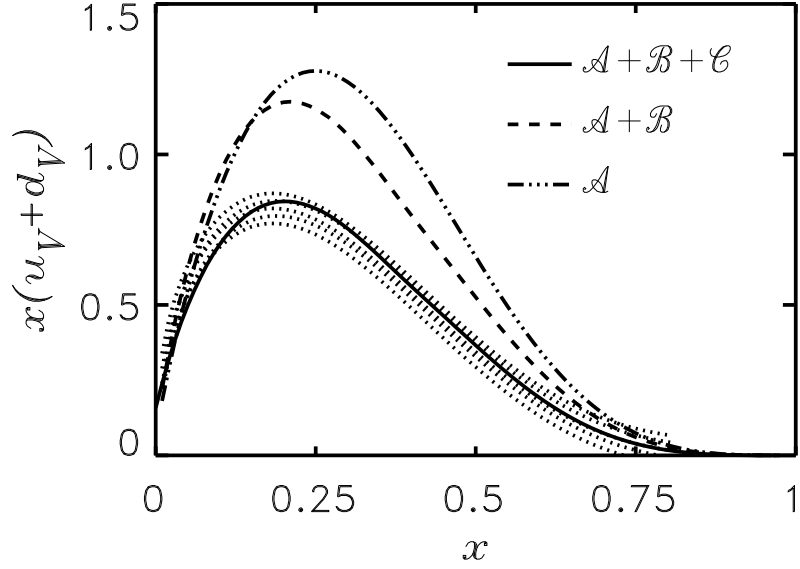,height=9cm}}
\caption{Total valence quark distribution evolved from
         $\mu^2 = 1$ GeV$^2$ to $Q^2 = 10$ GeV$^2$,
         from process ${\cal A}$ and ${\cal B}$, and ${\cal A}$ alone
         (dashed), and from ${\cal A}-{\cal C}$ together (solid).
         Shaded region represents data at $Q^2 = 10$ GeV$^2$.}
\label{F8}
\end{figure}

\newpage

\begin{figure}
\centering{\ \epsfig{figure=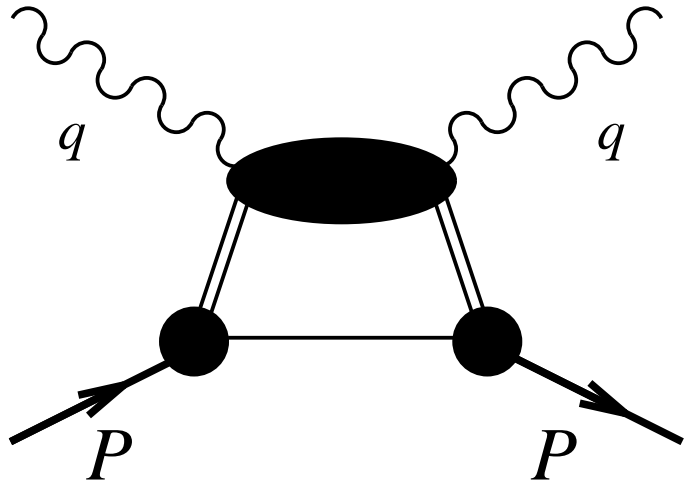,height=9cm}}
\caption{Deep-inelastic scattering from a diquark in the nucleon.}
\label{F9}
\end{figure}

\end{document}